\documentclass[%
reprint,
 amsmath,amssymb,
 aps,
]{revtex4-1}

\usepackage{graphicx}
\usepackage{dcolumn}
\usepackage{bm}
\usepackage{hyperref}

\usepackage{xcolor}
\usepackage{tabularx}
\usepackage{amssymb}

\begin{document}

\newcommand{\insertFigure}[3]{
\begin{figure}
  \includegraphics[width=\columnwidth]{#1}
  \caption{#3\label{#2}}
\end{figure}
}

\newcommand{\insertFigureFullPage}[3]{
\begin{figure*}
  \includegraphics[width=0.75\textwidth]{#1}
  \caption{#3\label{#2}}
\end{figure*}
}

\newcommand{\insertDoubleFigure}[4]{
\begin{figure*}
  \includegraphics[width=0.49\textwidth]{#1}
  \includegraphics[width=0.49\textwidth]{#2}
  \caption{#4\label{#3}}
\end{figure*}
}

\newcommand{\checkThis}[1]{\textcolor{red}{#1}}

\title{
Impact of a course transformation on students' reasoning  about  measurement uncertainty}

\author{Benjamin Pollard$^1$}
 
   \email{benjamin.pollard@colorado.edu}
\author{Alexandra Werth$^1$}
  \author{Robert Hobbs$^2$}

  \author{H. J. Lewandowski$^1$}
  \affiliation{$^1$Department of Physics, University of Colorado Boulder, Boulder, CO 80309, USA}
  \affiliation{JILA, National Institute of Standards and Technology, Boulder, CO 80309, USA }
    \affiliation{$^2$Department of Physics, Bellevue College, Bellevue, WA 98007, USA}

\date{\today}

\begin{abstract}
Physics lab courses are integral parts of an undergraduate physics education, and offer a variety of opportunities for learning.
Many of these opportunities center around a common learning goal in introductory physics lab courses: measurement uncertainty.
Accordingly, when the stand-alone introductory lab course at the University of Colorado Boulder (CU) was recently transformed, measurement uncertainty was the focus of a learning goal of that transformation.
The Physics Measurement Questionnaire (PMQ), a research-based assessment of student understanding around statistical measurement uncertainty, was used to measure the effectiveness of that transformation.
Here, we analyze student responses to the PMQ at the beginning and end of the CU course.
We also compare such responses from two semesters: one before and one after the transformation.
We present evidence that students in both semesters shifted their reasoning in ways aligned with the measurement uncertainty learning goal.
Furthermore, we show that more students in the transformed semester shifted in ways aligned with the learning goal, and that those students tended to communicate their reasoning with greater sophistication than students in the original course.
These findings provide evidence that even a traditional lab course can support valuable learning, and that transforming such a course to align with well-defined learning goals can result in even more effective learning experiences.
\end{abstract}

\maketitle


\section{Introduction}
Lab courses are an important part of physics undergraduate curricula \cite{olson2012engage,national2012discipline}.
These courses offer opportunities for learning that is critical to becoming a physicist in many different career paths.
For example, lab courses are natural settings for students to acquire experimental skills \cite{kozminski2014aapt,Heron2016}, practice scientific communication in a wide range of formats \cite{Moskovitz2011,Zwickl2013b,kozminski2014aapt}, and develop sophisticated beliefs and epistemologies around the nature of science \cite{Wilcox2018}; these goals are not often a primary focus in lecture or theory-focused courses \cite{kozminski2014aapt,Smith2020}.
Recent research has shown that there are lab courses that meet some of these learning goals \cite{Smith2020a, Pollard2020, Lewandowski2020,Lewandowski2018}, but that there is still room for improvement \cite{Holmes2017,Holmes2018}.
As such, an increasing number of lab educators are considering the variety of learning goals possible in lab courses, and working to align their courses to better achieve these goals.
Hand in hand, education researchers need to better understand the range of learning that occurs in lab courses, and identify teaching strategies that are effective at facilitating such learning.

\subsection{Measurement uncertainty as a learning goal}
In this work, we focus on measurement uncertainty as a learning goal of introductory physics lab courses.
Uncertainty analysis is a common learning goal in physics lab courses \cite{Fox2020}, thus the specifics of how it is taught are as varied as physics labs themselves.
Here, we highlight a few lab curricula that are discussed in literature that describe a focus on measurement uncertainty, as well as some research studies around learning of measurement uncertainty in labs.

The Scientific Community Laboratory (SCL), developed 
at the University of Maryland, centers around a series of research questions that aim to teach students how to produce, analyze, and evaluate scientific evidence \cite{Lippmann2003}. 
The SCL elevates measurement concepts to the same level of importance as physics concepts, recognizing them as critical for those broader skills.
In particular, the SCL focuses on sources of variation in data and the generalizability of results based on statistical significance, and also includes uncertainty considerations in experimental design \cite{Kung2005a}. 

The Student-Centered Activities for Large Enrollment Undergraduate Programs (SCALE-UP) Project at North Carolina State University includes labs with uncertainty considerations as learning goals \cite{Beichner2007}.
These activities focus on including uncertainties when reporting results, and using uncertainty when comparing data.
A test developed specifically for SCALE-UP further explored student ideas and approaches around measurement uncertainty \cite{Abbott2003}.

The Investigative Science Learning Environment (ISLE), developed at Rutgers University, includes inquiry activities that focus on sources of experimental uncertainty and ways to minimize them in the context of experimental design and iteration \cite{Etkina2007, Etkina2010}. 
ISLE integrates measurement uncertainty ideas, particularly around systematic uncertainties, in design of and reflection on laboratory experiments.

More recently, the Structured Quantitative Inquiry Lab (SQILab) at the University of British Columbia aims to teach measurement uncertainty in the context of critical thinking \cite{Holmes2014a}.
SQILab includes explicit instruction around skills and concepts related to distributions, and extends to the formalisims comparing results using statistical tests.
Research on SQILab also includes attitudes and beliefs related to measurement uncertainty \cite{Strubbe2016}.

Lastly, the physics labs that are part of the introductory calculus-based physics sequences at Cornell University have been transformed with measurement uncertainty as a learning goal.
The transformation frames its goals explicitly in the context of the AAPT lab guidelines \cite{kozminski2014aapt}, and identifies both statistical and systematic uncertainty learning outcomes in the context of modeling and experimental design \cite{Holmes2019}.
While the course included conceptual introductions to measurement uncertainty before it was transformed, research has shown that the more integrated and explicit approach in the transformed course resulted in more students viewing uncertainty as important when deciding if a result is trustworthy \cite{Smith2020b}.

\subsection{Measuring learning of measurement uncertainty}
In addition to developing physics lab curricula, physics education researchers have also studied student learning and student ideas around measurement uncertainty, often in conjunction with curriculum development \cite{Deardorff2001, LippmannKung2006,Holmes2014b,Majiet2018,Stein2020}.
Central to those efforts is the development of several research-based assessment tools related to measurement uncertainty \cite{Madsen2019, Pollard2017}.
The Concise Data Processing Assessment (CDPA) was developed around a decade ago to measure student understanding of both measurement uncertainty and mathematical models of measured data \cite{Day2011a}.
It has since been used to study pedagogical scaffolding \cite{Holmes2014} and gender differences in physics labs \cite{Day2016}.
Around the same time that the course transformation project at CU was initiated, the Laboratory Data Analysis Instrument (LDAI) was developed to measure data analysis skills within the context of a single lab report \cite{Eshach2016}.
While the LDAI does not focus on measurement uncertainty exclusively, it includes many aspects of measurement uncertainty as they relate to data analysis.
More recently, the Physics Lab Inventory of Critical Thinking (PLIC) was developed to measure a range of skills under the umbrella of critical thinking \cite{Walsh2019}.
Measurement uncertainty concepts are represented in the PLIC in the context of this broader range of experimental practice.

For this work, we use the Physics Measurement Questionnaire (PMQ) \cite{Campbell2005} to study the introductory lab course at the University of Colorado Boulder (CU), as both the course and the PMQ focus on statistical measurement uncertainty concepts at the introductory physics level. 
We first describe the course in Sections \ref{courseTrans} and \ref{MUtrans}.
We then describe the history and philosophical perspective of the PMQ in Section \ref{PMQsection}, and the particular items (or probes) of the PMQ on which this work focuses in Section \ref{PMQProbes}.

\section{Background}
\subsection{Transformation of an introductory lab course} \label{courseTrans}
In the broader context of improving physics lab education, the introductory lab course at the University of Colorado Boulder (CU) was recently transformed and studied.
We describe the course and the transformation process here; more details can be found in refs. \cite{Pollard2020,Lewandowski2020,Lewandowski2018,Pollard2018,Pollard2017,Lewandowski2017}.

The introductory physics lab course at CU is a stand-alone course typically taken by students in their second or third semester of study at CU.
For most students, it is the first physics lab course that they take at the college level.
The course, both before and after it was transformed, consists of a series of lab activities involving basic concepts from mechanics, electricity and magnetism, and other topics from introductory physics. 
Students meet weekly in two-hour lab sessions to work through each activity, and occasionally attend additional lecture sessions on background topics.
There are short pre-lab videos that students view before each activity, which include embedded questions for students to respond at particular points in each video \cite{Lewandowski2020}.
Students keep an electronic lab notebook while they work, which they upload for grading and feedback at the end of each activity.
The course has no midterms nor a final exam.

Beginning in 2016, author HJL began teaching the introductory physics lab course at CU.
At the same time, she initiated a project to transform this course.
First, professors in the Physics Department and various departments in the College of Engineering and Applied Science were surveyed, and engaged in group discussions, in order to identify learning goals for the course.
These goals included an alignment of students’ beliefs and epistemologies about experimental physics with that of expert physicists, positive attitudes about the course and about experimental physics more generally, 
the ability to create quality graphs, 
and an understanding of measurement uncertainty \cite{Lewandowski2018}.
Based on these learning goals, HJL, BP, RH, and others created a new set of lab activities for the course, with corresponding apparatus, analysis software, lab guides, grading rubrics, pre-lab videos, and lectures.
The transformed course was first taught in Fall 2018, and continues to the present.
HJL continued to teach the course throughout this process, including all the semesters studied below.

While the transformed course was designed to meet the identified learning goals, it is still distinct from the ideal course that the designers would have wished.
This mismatch is due mostly to logistical constraints such as those arising from working with 20-30 graduate teaching assistants, and the logistics of scheduling 35-45 separate weekly lab sections in a single instructional space. 
Thus, the transformed course still operates in many ways as a traditional introductory physics lab course.
For the context of this work, we see traditional lab courses as highly guided and prescriptive, focusing on conceptual rather than skills-based learning, and consisting of verification experiments. 
In particular for our transformed course, the lab activities remained quite prescriptive, guiding students through procedures with significant scaffolding throughout.
Nonetheless, most of the activities in the transformed course focused on skills-based learning, and none were verification experiments.

\subsection{Transformation aspects related to measurement uncertainty} \label{MUtrans}
In this work, we focus on the course transformation learning goal concerning measurement uncertainty.
The transformed introductory lab course at CU includes several aspects that support learning around measurement uncertainty.
First, each lab activity in the transformed course involves students measuring a quantity or outcome that they would not know before completing the measurement.
These activities are different than \textit{verification labs}, in which students are measuring a value that they learned in lecture or could look up in a textbook.
In the course before transformation, five out of six activities were verification labs, in our judgement.
In addition to there being no verification labs in the transformed course, many of the lab activities ask students to use measurements they made previously to make predictions about their present experiment.
Then, after making a measurement, many of the lab activities ask students to discuss their result with their peers in the classroom, comparing data to decide if their different results agree with each other.
These discussions provide repeated opportunities for students to consider and communicate both the value and the uncertainty of a result, and to discuss these results in the context of their choices involving data collection and procedure.

Beyond the lab activities themselves, four out of the six lectures in the transformed course focused entirely on measurement uncertainty concepts, and a fifth included additional discussion of measurement uncertainty along with other topics.
As these concepts are presented in the transformed course, students first learn about distributions and the act of measurement as sampling from a distribution.
The idea of uncertainty in measurement is presented as a measure of such underlying distributions.
While measurement uncertainty was included in lectures before the transformation, it was not as much a focus, and was presented with less of a conceptual underpinning, focusing more on the mechanisms of error propagation and the proper structures for reporting results.

\subsection{The Physics Measurement Questionnaire} \label{PMQsection}
The PMQ originated from studies by researchers in York, UK with primary school students age 9-16 \cite{Lubben1996,Volkwyn2005}.
This work stemmed from a need to evaluate a new national curriculum that included school laboratory programs \cite{Millar1994}, and resulted in a model for how students progressed in their ideas about measurement that categorized students' ideas concerning experimental data as a progression through eight levels. \cite{Millar1996}. Soon after, researchers in Cape Town, ZA attempted to use the materials from York in their physics classes for first-year university students at the University of Cape Town. 
They found, however, that the materials were not suitable in their context, so they created the PMQ by adapting the instruments developed in York \cite{Campbell2005}.

Similarly, as the Cape Town researchers interpreted preliminary responses from their students, they extended and adapted the frameworks from York to develop the point and set paradigms \cite{Buffler2001,Buffler2003}.
The point and set paradigms characterize two philosophical perspectives pertaining to the statistical uncertainty of measured quantities.
They are described in detail in refs. \cite{Volkwyn2008b,Campbell2005}. 

The point paradigm represents the idea that it would be possible for a single measurement trial to completely represent the true value of a physical quantity or measurand, where deviations from that true value are due to mistakes in the data taking procedure or unaccounted-for effects in the measurement apparatus.
The point paradigm would maintain that the overall goal of a good measurement procedure is to prevent, or identify and eliminate, all mistakes and unaccounted-for effects, allowing for a measurement trial that perfectly captures the measurand.
Thus, in the point paradigm, the results of individual trials can be considered independently of each other as long as all factors leading to deviation are taken into account. 

In contrast, the set paradigm represents the idea that each individual measurement trial reveals some information about the measurand, but that no individual measurement can yield its true value.
Thus, multiple trials must be considered as a distribution, with each successive trial revealing more information about the measurand.
However, perfect knowledge of the measurand with zero uncertainty is impossible under the set paradigm.
The set paradigm stems from a probabilistic approach to measurement uncertainty \cite{Buffler2003}, and is often considered to be more aligned with expert-like reasoning than the point paradigm.

It is also worth noting what is not captured by the point and set paradigms.
The paradigms, and by extension the PMQ itself, were designed to characterize reasoning around \textit{statistical} measurement uncertainty.
Discussions around systematic errors, that is, any unwanted or unaccounted-for effect that would not ``average out'' with repeated trials, are outside the scope of the point and set paradigms.
While students' responses in the PMQ often involve such reasoning, those elements are irrelevant in the framework of the point and set paradigms.
Additionally, skills and concepts concerning the propagation of uncertainty throughout a calculation are also outside the scope of the paradigms and the PMQ.
There are also more subtle distinctions to be made when discussing statistical measurement uncertainty, such as the differences between frequentest and Bayesian perspectives, that are more complex than the distinctions captured by the point and set paradigms.
Despite its limited scope, the PMQ is still a valuable tool for studying student learning around measurement uncertainty, especially at an introductory level.
Likewise, the point and set paradigms are nonetheless useful constructs for understanding overarching trends in learning regarding measurement uncertainty at the introductory physics level.

\subsection{This work}
In this work, we use the PMQ to measure the effectiveness of the introductory lab course at CU at facilitating student learning around statistical measurement uncertainty.
Such learning is directly related to one of our course transformation's learning goals.
In full, this goal was stated as, ``Students should demonstrate a set-like reasoning when evaluating measurements,'' where ``set-like'' is a reference to the set paradigm discussed in Section \ref{PMQsection}.

We aim to answer the research question, (Q1) \textit{Did students respond to the PMQ in ways more aligned with the set paradigm after taking the introductory lab course, compared to when they began the course?}
In answering (Q1), we consider both the original and transformed course, despite the fact that the original course did not have explicitly stated learning goals, to investigate whether an entirely traditional physics lab course can achieve such a learning outcome. 

Furthermore, we use the PMQ to evaluate the effectiveness of the transformation at achieving its learning goal around measurement uncertainty.
We aim to answer the research question, (Q2) \textit{Did student responses to the PMQ after the transformation show greater change towards the set paradigm than responses before the transformation?}

In Section \ref{MethodsSection}, we describe the probes of the PMQ that we use in this work, as well as the students who take the introductory physics lab course and our approach to collecting and analyzing responses from them. 
Section \ref{ResultsSection} presents results from our analysis of these PMQ responses, comparing responses from the start of the course (pre) and after completing the course (post), and from before the transformation and after it.
We finish by discussing these results in the broader context of physics lab courses in Sections \ref{DiscussionSection} and \ref{ConclusionsSection}.

\section{Methods} \label{MethodsSection}
\subsection{Probes of the PMQ} \label{PMQProbes}
The entirety of the PMQ is based on an experiment involving rolling a ball down a slope and then measuring the distance it travels in free fall.
Each item, or probe, of the PMQ concerns a decision at one step in the measurement process, from taking data to comparing the analyzed results.
Each item asks students to make a choice, usually between two or three multiple-choice options, and then to explain their choice in open-response format.

In this work, we analyze student responses to four particular probes of the PMQ: RD, UR, SMDS, and DMSS.
We chose to exclude the other probes of the PMQ from our analysis for a combination of reasons.
Some probes (RT, DMOS, and DMSU) only appear on the pre-test or the post-test version of the PMQ, but not on both, so it was not possible to directly compare students' responses to these probes between pre and post tests.
Additionally, when a fellow researcher in our group adapted the PMQ from the original pen-and-paper format to an online format using the Qualtrics online survey platform \cite{Qualtrics2005}, 
the SLG probe required significant adaption to the online format, so we decided not to code responses to that probe.
Finally, when we consulted with one of the creators of the PMQ, they recommended that we omit one of the probes (RDA) from our analysis, as they gained relatively little insight from that probe in their work.

As an example, the RD probe is shown in Fig. \ref{RDProbe}.
The RD probe concerns data collection.
RD stands for ``repeated distance,'' referring to whether to repeat a trial that measures the distance that a ball travels.
The probe presents three stances: to repeat the trial several more times, to move on after performing only a single trial, or to repeat the trial exactly one more time.
Respondents are asked to choose with  which stance they agree, and to explain their choice in a text box.

\insertFigureFullPage{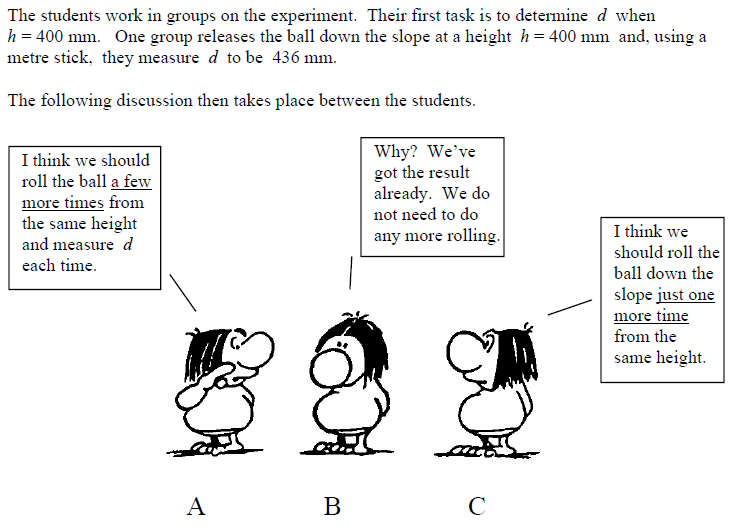}{RDProbe}{The RD probe of the PMQ. Respondents are asked, ``With whom do you \textbf{most closely} agree?'' and can choose A, B, or C. They are then prompted to ``Explain your choice.'' Reproduced from ref. \cite{Volkwyn2008b}.}

The other probes of the PMQ have the same general format as RD, and are shown in Appendix \ref{AppProbes}.
The UR probe, which stands for ``using repeats,'' asks how to analyze data to produce a final result.
The SMDS probe, which stands for ``same mean, different spread,'' concerns data comparison, asking respondents to decide which of two data sets is better.
The two data sets have the same mean, but different spread.
Similarly, the DMSS probe, which stands for ``different mean, same spread,'' also concerns data comparison.
However, the DMSS probe concerns two data sets that have different means but the same spread.

\subsection{Coding scheme development}
Because every probe in the PMQ involves an open-response component, responses must first be classified qualitatively before additional analysis is performed.
To classify PMQ responses, the creators of the PMQ developed a coding scheme based on 
responses from their students \cite{Volkwyn2005}.
The coding scheme consists of a different set of codes for each probe, and aims to capture the types of reasoning students draw upon when reasoning about uncertainty in the various stages of measurement in the PMQ.
This coding scheme was developed concurrently with the paradigm model described above, and in the current version each code is associated with either the point paradigm, the set paradigm, or an ``unknown'' designation if the code represents reasoning that does not unambiguously align with one paradigm.

The first set of PMQ responses that we analyzed came from students in the course in Fall 2016, before any of the responses that we present here.
We observed a range of student reasoning in the 2016 data set that was not captured by the coding scheme developed by the creators of the PMQ, and thus expanded upon it to describe responses from our different national, institutional, and course context. 
These expanded codes were subsequently consolidated into common themes, which were then re-framed as code definitions to create a new coding scheme for the PMQ.
This new coding scheme was then refined based on CU student responses from Fall 2017, with a subset of those responses used to check inter-rater reliability between two independent coders.
The process of creating and refining the new coding scheme is described in more detail in ref. \cite{Pollard2020}.

After our coding scheme was developed and verified, we applied it to data from the Spring 2017 and Spring 2018 semesters.
We first matched pre and post responses by student, and removed the responses that were not matched from the data set to make direct comparisons of pre and post distributions straightforward.
Then, for each probe, all of the pre- and post-test responses from those two semesters were anonymized, combined, and shuffled into a single data set, which RH coded without knowing from which semester and pre/post designation each response came.
After the codes were assigned, the data were separated back into their respective categories for further analysis.

\subsection{Coding scheme} \label{RDCodesDiscussion}
The new coding scheme we created consists of 12-16 codes per probe.
Each code is denoted by a letter and a number, for example, ``U3.''
When it is necessary to disambiguate between codes pertaining to different probes, we prepend the probe's acronym and a hyphen, for example, ``RD-U3.''

The letter is either S, P, or U, signifying whether the code falls under the set paradigm, the point paradigm, or unknown reasoning that does not unambiguously align with one paradigm.
Across all probes, there are 21 P codes, 22 S codes, and 13 U codes.
The codes within each paradigm further differentiate between student reasoning at a finer-grained level.
The number designation of each code distinguishes between them, though we do not intend for the numbers to be interpreted as an ordering.
The relative merits of each code are not inherent to the reasoning they represent, and in practice will depend on the context of how the results of analysis are interpreted.
For example, when we use the coding scheme here to measure the success of a course, we compare codes in terms of the extent to which each code's reasoning aligns with the learning goals of the course.

Sometimes, student responses contain multiple distinct lines of reasoning, and thus we allow multiple codes to be assigned to a single response.
In the data sets analyzed below, 87.9\% of the responses were assigned a single code, 11.9\% were assigned two codes, and 0.3\% were assigned three codes.
For the purposes of classifying a response into a single paradigm, if a response was assigned multiple codes from different paradigms, S and P codes both took precedence over U codes.
For example, if a response was assigned a P code and a U code, the response was considered point-like overall.
If both an S code and a P code were assigned to a single response, which happened in 1.7\% of the responses in the data set analyzed below, we classify that response's paradigm as U. 

The complete code books of the new coding scheme, one for each of the four probes we analyzed, are reproduced in full in the supplemental information accompanying this work.
Here, we present an subset of these codes in Table \ref{codeTable}.
As an illustration of the scope and depth of the coding scheme, we discuss here three codes for the RD probe: S4, P2, and U1.
Each code represents a reason to perform more than one trial, aligned with response A or C of the RD probe (Fig. \ref{RDProbe}).

The S4 code argues that multiple measurements should be performed in order to reduce the uncertainty of a mean value, implying that the mean is the result that matters.
This argument aligns with the set paradigm.
On the other hand, the P2 code represents the idea that multiple measurements are beneficial because they allow the experimenter to identify outliers or mistakes in data collection.
This argument aligns with the point paradigm.

Lastly, the U1 code represents responses that merely state that more data is needed.
In this case, the respondent did not write a sufficient explanation to classify it into one paradigm or the other.
It is possible that, if discussing the probe with the respondent in person, their underlying reasoning would become apparent.
It is also possible that the respondent lacked the language to express their reasoning, or that they had not considered their reasoning to any greater depth.
It is even possible that the student was merely pressed for time when completing the survey, and otherwise they would have provided an explanation that aligned well with another code.
In any case, the PMQ coder has only the written response to interpret, and as such, is forced to assign a code such as U1 regardless of these hypothetical cases.

We note that, from the perspective of an expert experimental physicist, there is validity behind the reasoning represented by both the S4 and the P2 codes, and even the U1 code cannot be said to be incorrect.
Therefore, we do not intend to assign an inherent ranking or hierarchy to the codes in our code book.
They aim only to classify common lines of reasoning used when responding to a given PMQ probe.
Even at the paradigm level, though the set paradigm has been identified as more expert-like than the point paradigm in previous work, that does not mean that point-like reasoning is not sometimes included in expert approaches more generally.
For example, outliers are often the impetus for proposing causes and enacting revisions in the larger context of modeling, and identifying systematic errors is sometimes merely a more nuanced framing for catching mistakes in the measurement procedure \cite{Dounas-Frazer2018b}.
When using our coding scheme to measure the effectiveness of a learning experience, as we do in this work, we intend for our codes to be compared to each other only in terms of how well they align with the goals of that learning experience.
In general, interpreting results from our coding scheme by ranking codes relative to each other is best done in the context of a learning goal.

\begin{table*}[htbp]
  \caption{Selected codes from the new PMQ coding scheme.\label{codeTable}}
  \begin{ruledtabular}
    \begin{tabular}{p{.05\linewidth} p{.07\linewidth} p{.26\linewidth} p{.6\linewidth}}
     Probe & Identifier & Name & Definition: "Argument is that..." \\ 
      \hline
      RD & S4 & Reduce uncertainty of mean & ...multiple measurements will be used to reduce the error/uncertainty of the mean/average.  \\
      RD & P1 & Measure the true value & ...the experimenter could measure the correct value in a single measurement. \\
      RD & P2 & Identify the outliers after all measurements & ...repeated measurements are needed in order to know which measurements were mistakes or outliers, after all measurements are taken. This code includes the idea that the experimenter must get the same result at least twice for it to be correct.\\
      RD & U1 & Just take more data & ...experimenter needs to take more data. No statistical reasoning apparent.\\
      UR & S1 & Simply “average” & ...“I averaged,” “do the average,” “average is best,” or “it is the average,” but does not elaborate. Includes statements that simply say what the reported value is.\\
      UR & S3 & Why average is appropriate in this case & ...reporting the average is best because all of this data matters, or because the spread of this data is small enough. Includes reporting all data as well as the average. \\
      UR & S4 & Report average and spread & ...experimenter should report the average and the uncertainty/range/spread.\\
      UR & S5 & How to compute & ...how to compute the average.  \\
      UR & P1 & Choose single value &...experimenter should choose a single value to report (for any reason). \\
      SMDS & S2 & Smaller spread is better, no mention of external factors & ...a smaller spread/uncertainty/range is better / more accurate / more precise / etc. The response does not mention external factors, outliers, human error, etc.\\
      SMDS & S3 & Smaller spread is better, due to external factors & ...a smaller spread/uncertainty/range is better / more accurate / more precise / etc. The response mentions external factors, outliers, human error, etc.\\
      SMDS & P1 & The means are the same & ...the groups agree because the means are the same.\\
      SMDS & P4 & Differences in carefulness & ...differences in the spread are due to differences in how carefully the measurements were performed. \\
      DMSS & S3 & Similar means and spreads, mentions overlap & ...the groups agree because the means and spreads are similar. Argument considers the overlap between the means and/or spreads of the two datasets.\\
      DMSS & S4 & Chose A, blank explanation & Respondent chose "A" but left the explanation blank.\\
      DMSS & P2 & Means must match & ...the groups do not agree because the means are not the same (no mention of spread)\\
      DMSS & P3 & Means close enough, treats average as point & ...the groups agree because the means are close enough\\
      DMSS & U1 & Not about statistics & ...only non-statistical things, such as systematics, are mentioned.\\
      DMSS & U3 & Misc. & Argument that doesn't fit into any of the other codes.\\
    \end{tabular}
  \end{ruledtabular}
\end{table*}

\subsection{Student-level paradigms}
In addition to interpreting the codes assigned to each students' response to any particular probe, we also consolidate students' responses to each of the four probes we analyzed into a single designation of that student's reasoning overall \cite{Pollard2017}.
Because of the coarse-grained nature of this consolidation, we characterize student's reasoning only at the paradigm level.
These student-level paradigms correspond to the number of probe-level paradigms emerging from each probe, as defined in Table \ref{studentParadigms}.
A student's overall response is \textit{point-like} if their responses to the probes are represented only by P and U codes and no S codes.
Conversely, an overall \textit{set-like} response comes from responses to the probes that are represented only by S and U codes and no P codes.
A third designation, \textit{mixed}, refers to an overall response that includes both P and S codes, or to when the response is entirely represented by U codes.
However, a response represented entirely by U codes only occurred 0.2\% of the time in the data sets analyzed here.

\begin{table}[]
\centering
\caption{Definitions of overall student paradigms. Reproduced from ref. \cite{Pollard2017}}
\label{studentParadigms}
\begin{ruledtabular}
\begin{tabular}{l c c}
\multicolumn{1}{l}{Student paradigm}     & \multicolumn{1}{c}{Number of \textit{P}'s}  & \multicolumn{1}{c}{Number of \textit{S}'s}   \\ \hline \noalign{\smallskip}
\textit{point-like} & $\geq1$						 & 0       						\\ 
\textit{set-like}   & 0       						 & $\geq1$ 						\\  
\textit{mixed}    & $\geq1$ 						 & $\geq1$ 						\\  
\textit{mixed}    & 0        					 & 0       						\\  
\end{tabular}
\end{ruledtabular}
\end{table}

\subsection{Statistical methods}
After responses were coded using the coding scheme described above, we analyzed the distributions of the assigned codes.
Note that in this work, we compare distributions of students within a semester and PMQ administration (pre or post), rather than comparing matched responses student-by-student. 
We compare distributions of codes and demographic characteristics using Fisher's Exact test \cite{Fisher1922}, with a significance threshold of p$<$0.05.
When considering the significance of multiple tests at once, we apply the Holm-Bonferroni method \cite{Holm1979} to correct for the problem of $m$ multiple comparisons. 
We compute these tests using the base package included in the R programming language, version 3.6.2 \cite{RCoreTeam2017}.
For an additional visual indication of the uncertainty in counts or percentages of codes or paradigms, we plot the binomial proportion confidence interval at the 95\% confidence level. 
When plotting the difference between the number of post responses and pre responses for each code, we propagate these confidence intervals for the plotted uncertainty bar of the calculated difference.
Finally, when two similarly-measured proportions are both statistically significant based on these methods, we estimate the degree to which they are different by calculating an effect size using Cohen's h \cite{cohen2013statistical}. 

\subsection{Course context}
In this work, we compare two semesters of the introductory physics lab course at CU, one before the transformation of that course (Spring 2017, the ``original course'') and one after the course was transformed (Spring 2018, the ``transformed course'').
We compare two Spring semesters, though the course is also taught in Fall semesters, to avoid a range of factors that influence students of differing backgrounds enrolling in the Fall versus in the Spring.
There were 641 students who completed the course at CU in Spring 2017, and 722 in Spring 2018.
Of these students, 539 and 499 respectively completed both the pre-test and post-test, and were included in the data set analyzed here.
The self-reported gender, race and/or ethnicity, major, and year of students in these two semesters, collected using another research-based assessment that was administered at the same times as the PMQ, are shown in Table \ref{demoTable}.
We include this information for various reasons \cite{AmyNoelleParks2012}, including to provide context for our research findings, as well as to enable meta-studies that combat normative whiteness and highlight inequities in research \cite{Kanim2020}.
We compared the proportions of students identifying with each of these categories in Spring 2017 and Spring 2018 as an indication of the similarity of the students entering the course during these two semesters.
The resulting p-values from Fisher's Exact are shown by each category heading in Table \ref{demoTable}.
Along each of these dimensions, the populations of students in the two courses were statistically equivalent (p$>$0.05).

\begin{table}[htbp]
\caption{Self-reported gender, race/ethnicity, major, and year of students enrolled in the course in both Spring 2017 and Spring 2018. ``Engineering'' excludes the major Engineering Physics, which is included in ``Physics.'' The p-values from Fisher's Exact comparing the two semesters appear next to each dimension heading. \label{demoTable} }
\newcolumntype{R}{>{\raggedleft\arraybackslash}X}
\begin{tabularx}{\columnwidth}{l|R}
\hline \hline 
\textbf{Gender} & p=0.31 \\
Female & 23.6\% \\
Male & 75.1\% \\
Other Gender & 1.3\% \\ 
\hline
\textbf{Race / Ethnicity} & p=0.81 \\
American Indian or Alaska Native & 0.9\% \\
Asian & 14.4\% \\ 
Black or African American & 2.2\% \\
Hispanic/Latino & 8.8\% \\
Native Hawaiian or other Pacific Islander & 0.7\% \\
White & 69.0\% \\
Other race/ethnicity & 4.0\% \\
\hline
\textbf{Major} & p=0.48 \\
Physics & 17.2\% \\
Engineering & 44.8\% \\
Other STEM & 35.1\% \\
Other disciplines & 3.0\% \\
\hline
\textbf{Year} & p=0.30 \\
First year & 48.9\% \\
Second year & 31.5\% \\
Third Year & 11.4\% \\
Fourth year & 6.2\% \\
Fifth year and above & 2.0\% \\
\hline \hline
\end{tabularx}
\end{table}

The PMQ was administered electronically to students at the beginning (pre) and at the end (post) of the course during both semesters.
Students were sent an internet link to complete the PMQ independently, and as an incentive for completing the questionnaire, were offered a small amount of participation-based course credit totaling 1-2\% of their final grade in the course.
In the original course, students completed the pre-test and post-test as in-class assignments.
However, due to unavoidable scheduling circumstances, students in the transformed course completed the surveys outside of class.
In that semester, students received the pre-test link five days after the course started, and were required to complete their responses within seven days.
Previous research on other online research-based assessments of student learning has shown little difference in matched student responses when completed outside or during class time, showing at most a small positive increase from taking the assessment in-class overall \cite{Wilcox2016a}. 

Additionally, because of the change in timeline in the transformed course, the first lecture in that course occurred before 64\% of the respondents completed their pre-survey response.
The content of that lecture touched on aspects related to measurement uncertainty, specifically the importance of every measurement having an associated uncertainty and how that uncertainty is used for comparing measurements.
For the post-tests, students received the link close to the end of the semester, after they had completed all activities for the course, and were required to complete their response before the semester ended.

\section{Results} \label{ResultsSection}
Here, we present quantitative results from applying the PMQ coding scheme developed at CU to two semesters of the introductory physics lab course at CU: Spring 2017 (referred to as the original course, or before the transformation) and Spring 2018 (referred to as the transformed course, or after the transformation).
For each semester, we compare the distribution of responses from the pre-test to equivalent distributions from the post-test, as a ``pre-to-post'' comparison. 
We first present these comparisons at the student paradigm level, as the most simplified interpretation of our results.
We then break down the results probe by probe, still interpreting responses at the paradigm level.
Lastly, we consider each probe at a level beyond the paradigms, comparing distributions of the codes that make up the paradigms.
At each level, we note how each finding aligns or runs counter to the learning goal of the transformed course.
We further discuss this alignment more broadly in the context of our research questions in Section \ref{DiscussionSection}.

\subsection{Paradigm-level results}
Fig. \ref{StudentParadigms} shows the percentage of students whose responses to the PMQ fell into each student-level paradigm, \textit{point-like}, \textit{mixed}, or \textit{set-like}.
The left panel shows the semester before the transformation, while the right show the semester after the transformation.
Light gray bars represent the pre-test, while darker gray represent the post-test.

The error bars in Fig. \ref{studentParadigms} suggest that, for both semesters, there were significant differences between pre and post for the number of \textit{mixed} and the number of \textit{set-like} responses. 
Fisher's Exact confirms those differences, all with $p\ll0.05$.
However, the proportions of \textit{point-like} responses were statistically similar ($p=1$ for 2017 and $p=0.34$ for 2018).
Overall at the student level, students shifted predominantly from the \textit{mixed} to \textit{set-like} paradigm, both before and after the transformation.
\insertDoubleFigure{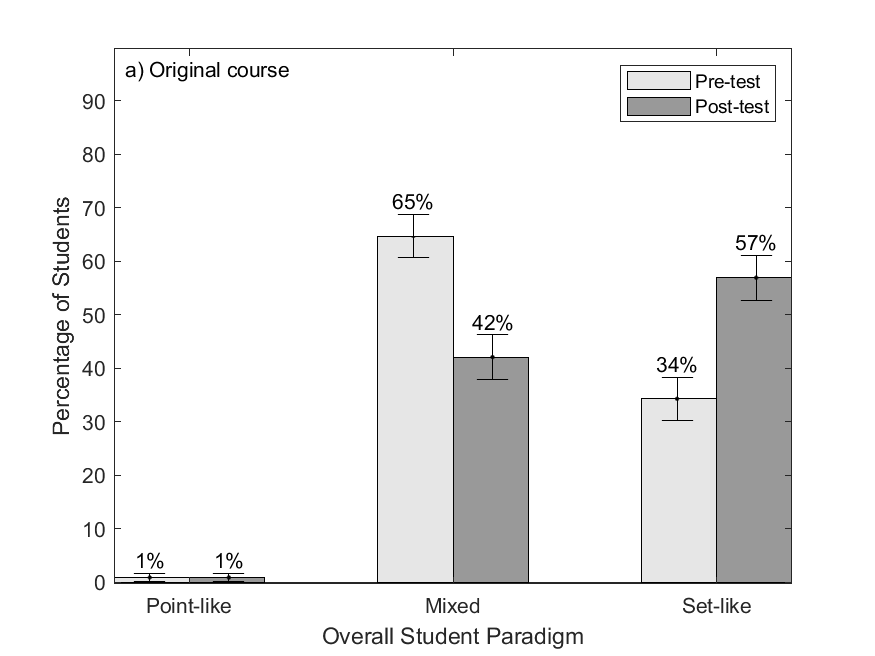}{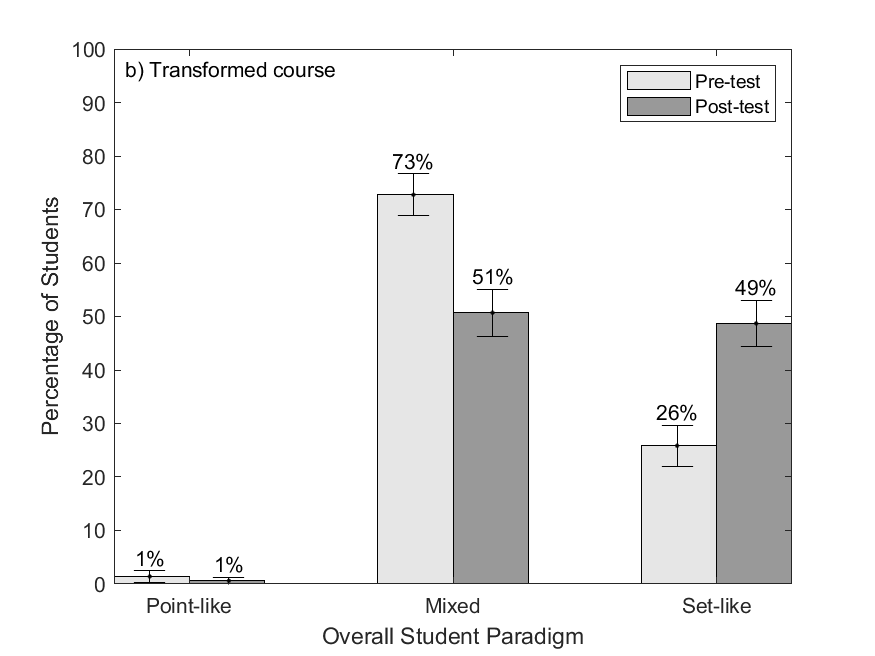}
{StudentParadigms}{Pre-post shifts at the student paradigm level. (a) is before transformation, (b) is after transformation. Error bars are the binomial proportion confidence interval at the 95\% confidence level.}

Moving to the probe level, shifts between pre and post paradigms for each of the four probes we analyzed are shown in Fig. \ref{ProbeParadigms}.
As before, the left panel shows the semester before the transformation, while the right show the semester after the transformation.
Within each panel, the four probes are represented on the vertical axis.
The horizontal axis represents the proportion of students whose responses were coded with either $S$ or $P$ codes.
Solid markers denote the proportion with an $S$ code on the pre-test, while open markers denote the proportion with a $P$ code on the pre-test.
The end of the corresponding arrows show the corresponding proprtions on the post-test.
The significance of these shifts is also indicated.

Across all probes, students shifted to more $S$ reasoning and less $P$ reasoning during the semester.
However, only some of these shifts were statistically significant.
In the RD probe, the transformed course showed larger shifts towards $S$ than the original course (with Cohen's h effect sizes of h=0.45 and h=0.21, respectively), while in the SMDS probe, the transformed course showed no significant shifts at all.
The UR probe showed little practical significance in either semester due to the large proportion of $S$ responses, even on the pre-test.
We have seen this saturation effect before in the UR probe when analyzing responses from CU students at the paradigm level \cite{Lewandowski2017,Pollard2020}.

The DMSS probe shows a larger shift towards $S$ reasoning in the transformed course than in the original course.
However, this shift is due to a difference in the proportion of $S$ responses in the pre-tests, rather than the post-tests.
Such a difference stands in contrast to the other three probes, which show similar pre-test proportions between the two semesters.
Furthermore, this difference in pre-test responses stands in contrast to all of the information we have available on the distribution of students who enrolled in the class for these two semesters, which would suggest that the two groups of students are similar. 
Our best guess as to the cause of this difference in the DMSS pre-test proportions concerns the differences in the timelines of the two semesters.
We speculate that responses in the transformed semester to the DMSS probe in particular were affected by the first lecture of the course, for the 64\% of students who completed the survey after that lecture.
That lecture touched on the idea that uncertainty is used for comparing measurements in a generalized way, an idea that relates to the DMSS probe.
However, it also relates to the SMDS probe, from which the paradigms of pre responses seem similar before and after the transformation. 
Given this uncertainty, we proceed with caution when further analyzing DMSS responses from these two semesters, remembering that the full story around this portion of the data set remains unclear. 

\insertDoubleFigure{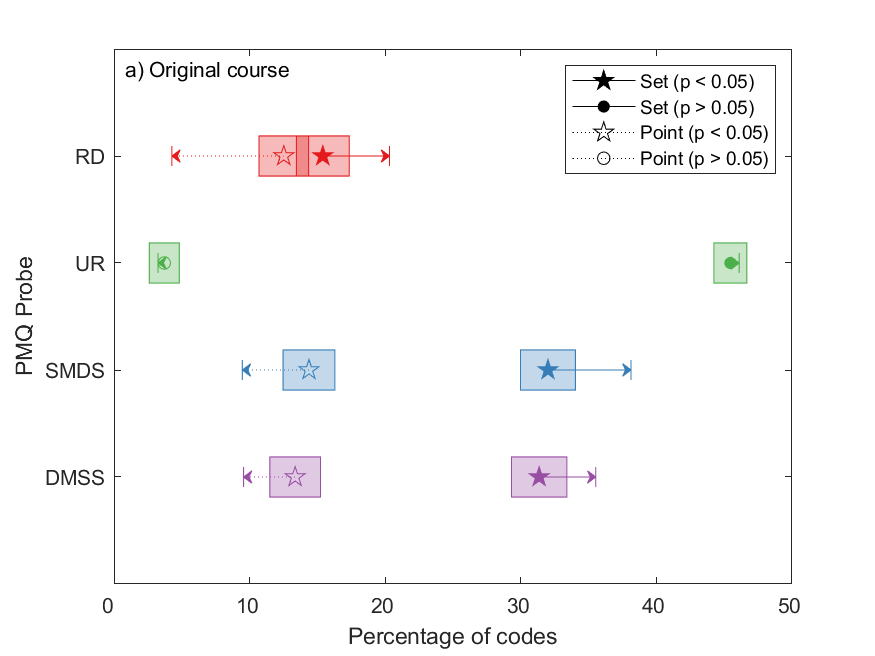}{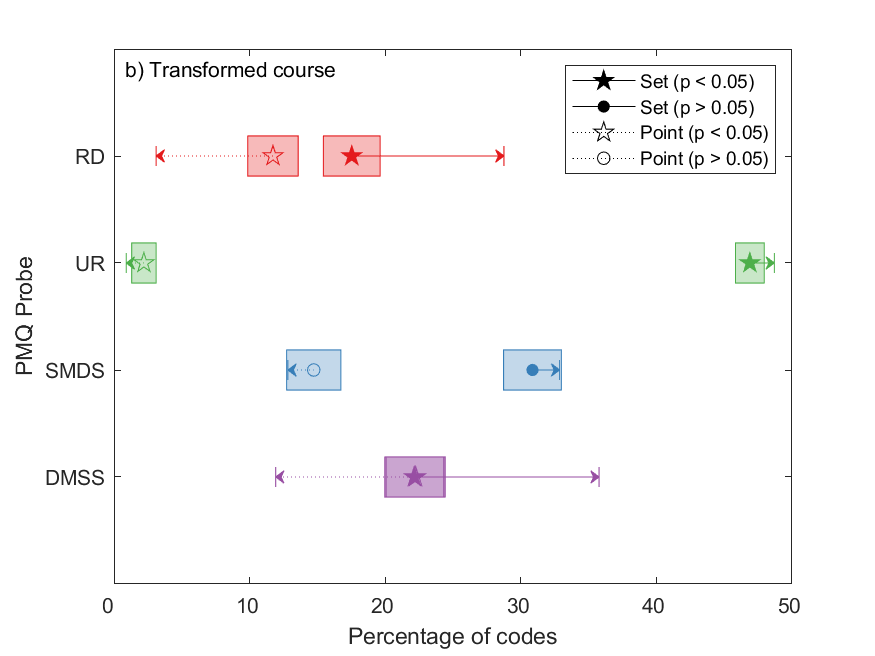}
{ProbeParadigms}{Pre-post shifts at the probe paradigm level. (a) is before transformation, (b) is after transformation. The horizontal axis represents the proportion of students responding in either the point or set paradigm, on each of four probes along the vertical axis. The shapes at the start of each arrow represent the pre-test proportion, while the location of the arrowhead represent the post-test proportion. Solid shapes represent the set paradigm proportion, while open shapes represent the point paradigm proportion. Star shapes represent a statistically significant pre-to-post shift using Fisher's Exact test at the 95\% confidence level; circle shapes represent a shift that is not significant by that same test. Shaded boxes represent the pretest binomial proportion confidence interval at the 95\% confidence level, as a guide to the eye.}

\subsection{Code-level results}
We now analyze results beyond the level of paradigms, considering the individual codes themselves in the context of the course transformation.
For each probe and semester, we plot the difference between the number of responses to each code on the post-test and on the pre-test.
In addition to the error bars that represent the uncertainty of these differences, we use blue bars to represent codes in which the pre- and post-test distributions are statistically different using Fisher's Exact ($p<0.05$), and yellow bars for the codes in which the pre- and post-test distributions are statistically the same ($p>0.05$).
We apply the Holm-Bonferroni correction, with $m$ as the number of codes in the given probe's code book, to the $p$-values from Fisher's Exact before determining statistical significance.

\subsubsection{The RD probe}
\insertDoubleFigure{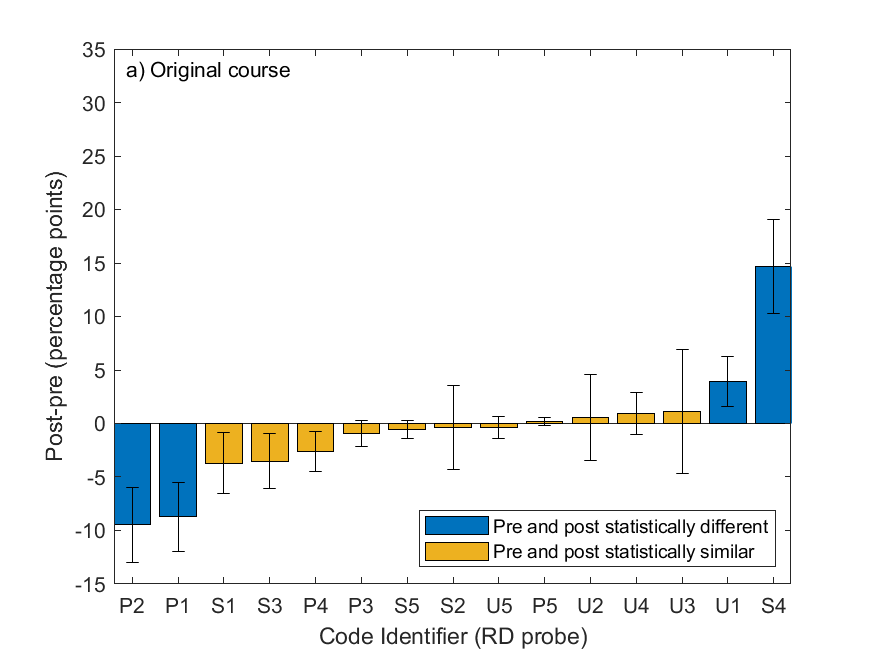}{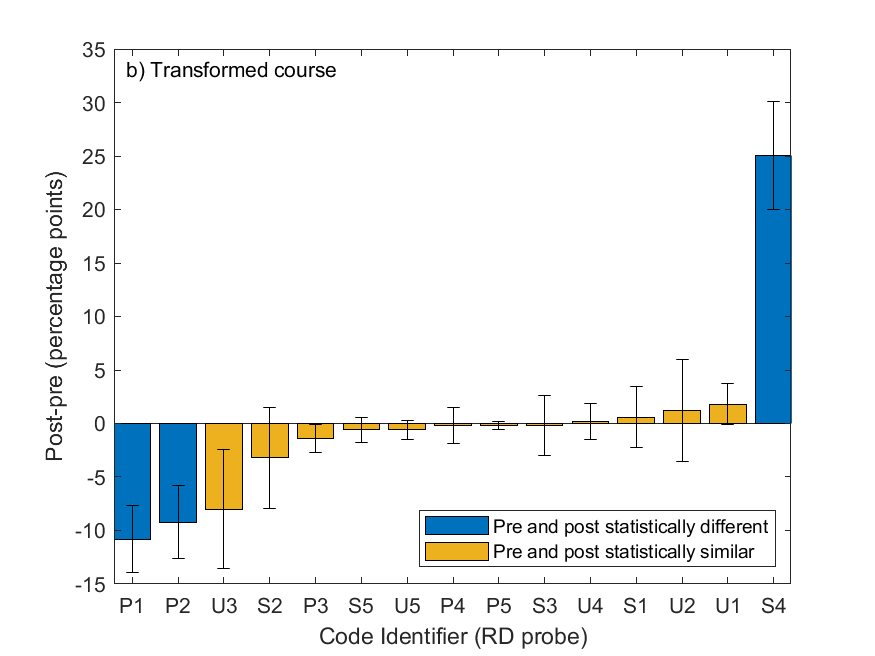}
{RDCodes}{Pre-post differences in code counts for the RD probe, for the original course (a) and the transformed course (b). Blue bars are statistically significant differences using Fisher's Exact test at the 95\% confidence interval, adjusted using the Holm-Bonferroni method. Organge bars are are not significant by that same test. Error bars are the binomial proportion confidence interval at the 95\% confidence level.}
A comparison for the codes in the RD code book is shown in Fig. \ref{RDCodes}.
In both semesters, the code with the largest change from pre to post was S4.
The prominent increase in S4 is encouraging, as it is aligned with the learning goals of the course, in particular the idea that all numbers have an uncertainty, including the mean of a set of data.
Before the transformation, the other code that increased was U1, while that code did not significantly change after the transformation.
U1 represents a response that does not display more sophisticate reasoning than the other codes.
Therefore, the lack of an increase in U1 after the transformation compared to before could suggest that students articulated their reasoning with greater sophistication in the transformed course.

Considering the codes that decreased from pre to post, there were two codes that showed a significant pre-to-post decrease, and they were the same codes in both the transformed and the original course.
As they were both point-like, that change is aligned with the goals of the transformation.

\subsubsection{The UR probe}
\insertDoubleFigure{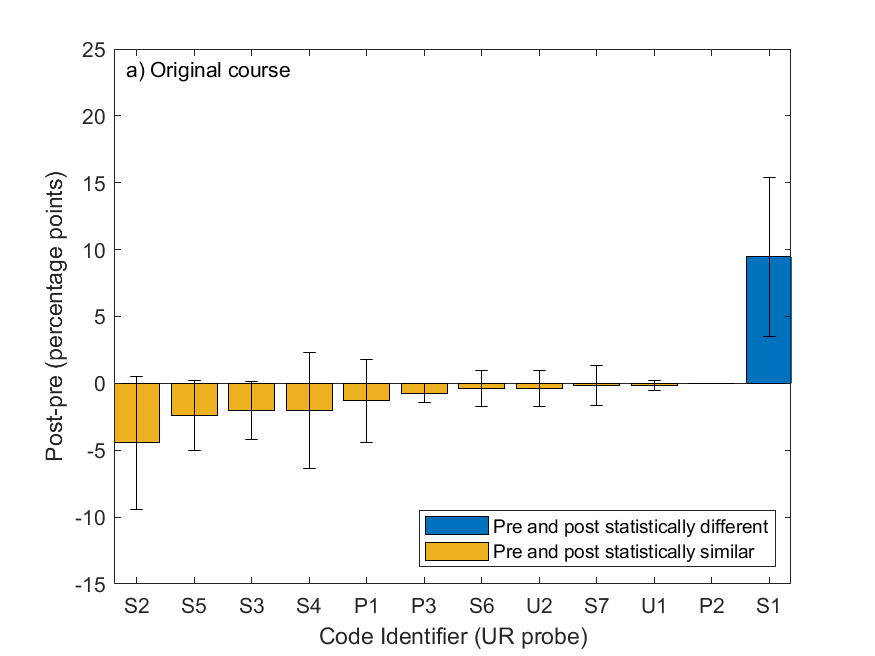}{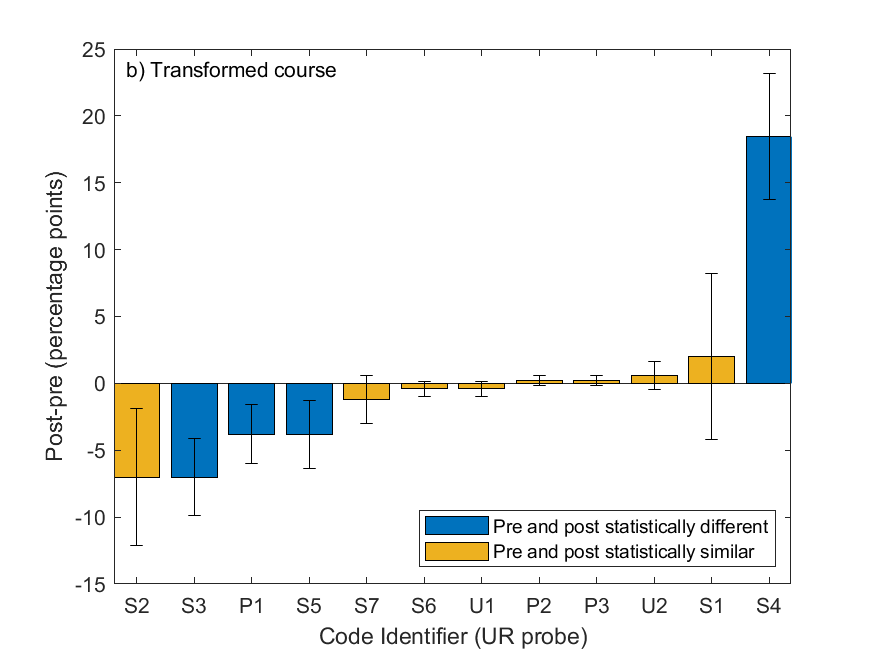}
{URCodes}{Pre-post differences in code counts for the UR probe, for the original course (a) and the transformed course (b). (b) reproduced from ref. \cite{Pollard2020}.  Blue bars are statistically significant differences using Fisher's Exact test at the 95\% confidence interval, adjusted using the Holm-Bonferroni method. Organge bars are are not significant by that same test. Error bars are the binomial proportion confidence interval at the 95\% confidence level.}
Fig. \ref{URCodes} shows a comparison of the codes in the UR code book.
In both semesters, the most prominent change was an increase of a single set-like code.
This consolidation phenomenon in UR responses is discussed in ref. \cite{Pollard2020}.
In the original course, the code into which students consolidated was S1, while in the transformed course the code as S4.
S1 represents reporting the average as the result of a set of measurements, while S4 represents reporting an average as well as a spread.
S4 aligns with the transformation's learning goals, as it recognizes the importance of the spread of a distribution and aligns with the idea that all numbers have an uncertainty.

There were no other statistically significant pre-post decreases in UR responses in the original course.
However, there were three significant decreases in the transformed course: P1, S3, and S5.
P1 represents canonical point-like reasoning, that one should choose the value from a single trial to represent the result of an experiment.
A decrease in this code is aligned with the goals of the transformation.
The other two codes that decreased in the transformed course were set-like.
One of them, S3, discusses the purpose of reporting an average, representing conceptual reasoning around the role of averages in experimentation.
The second, S5, states the mathematical process of calculating an average, and represents the basic skill or practice of reporting an average.
The transformation's goals included  
the reasoning represented by both of these codes,
suggesting that the decreases in S3 and S5 may represent further room for improvement.

\subsubsection{The SMDS probe}
\insertDoubleFigure{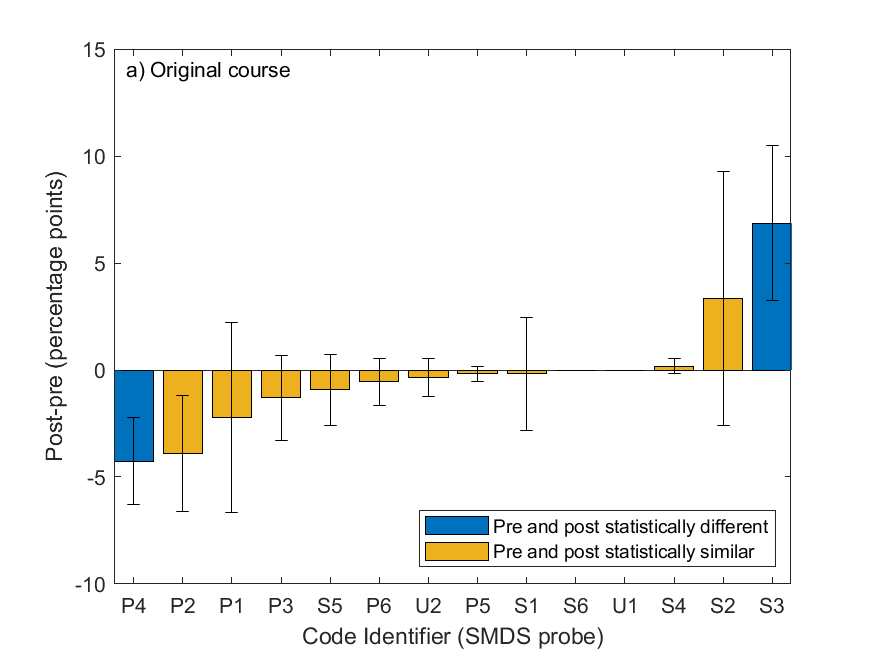}{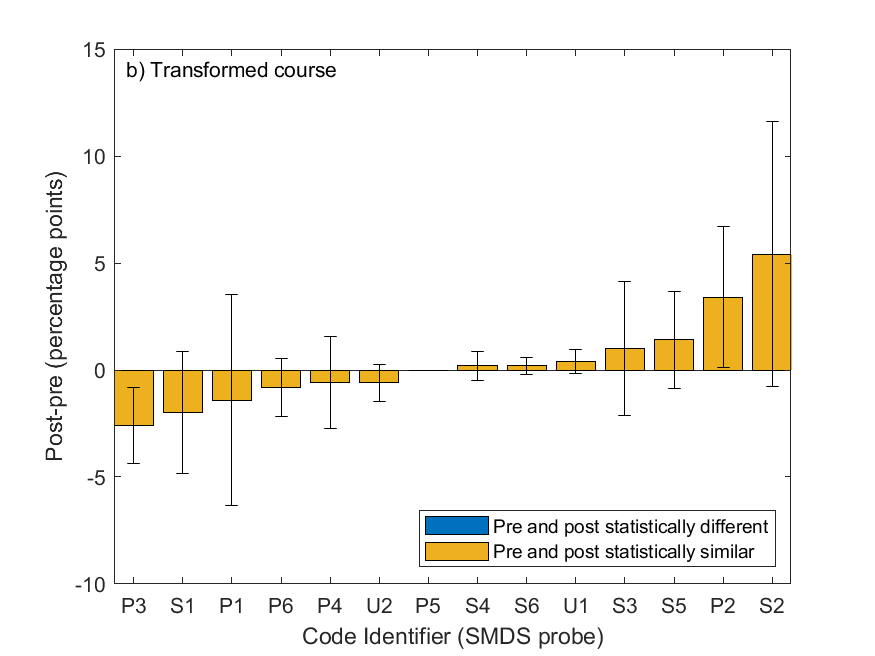}
{SMDSCodes}{Pre-post differences in code counts for the SMDS probe, for the original course (a) and the transformed course (b). Blue bars are statistically significant differences using Fisher's Exact test at the 95\% confidence interval, adjusted using the Holm-Bonferroni method. Organge bars are are not significant by that same test. Error bars are the binomial proportion confidence interval at the 95\% confidence level..}
We compare responses to the SMDS probe in Fig. \ref{SMDSCodes}.
Overall, the magnitude of the shifts on any of these codes are relatively small, suggesting that the type of reasoning solicited by the SMDS probe is relatively stable in our population of students.
In fact, in the transformed course, no single probe showed a statistically significant change pre-to-post.

Before the transformation, the code that significantly increased was S3, which states that a smaller spread is better because of external factors such as ``air resistance'' or ``human error.''
While the recognition of spread playing a role in data comparison is aligned with set-like reasoning, and thereby the goals of the transformation, the focus on external factors over inherent statistical variation is more point-like than set-like.
The code that decreased pre-to-post before the transformation was P4, which talks about differences in carefulness between the two experimenters.
This idea aligns with the point paradigm if the lack of carefulness manifests as mistakes in individual trials.
However, there is a subtle difference between this line of reasoning and the idea that the spread of a data set overall is affected by differing tendencies of experimenter care.
Taken together, the SMDS trends observed in the original course contain elements that are 
both closer and further away from set-like reasoning.


\subsubsection{The DMSS probe}
\insertDoubleFigure{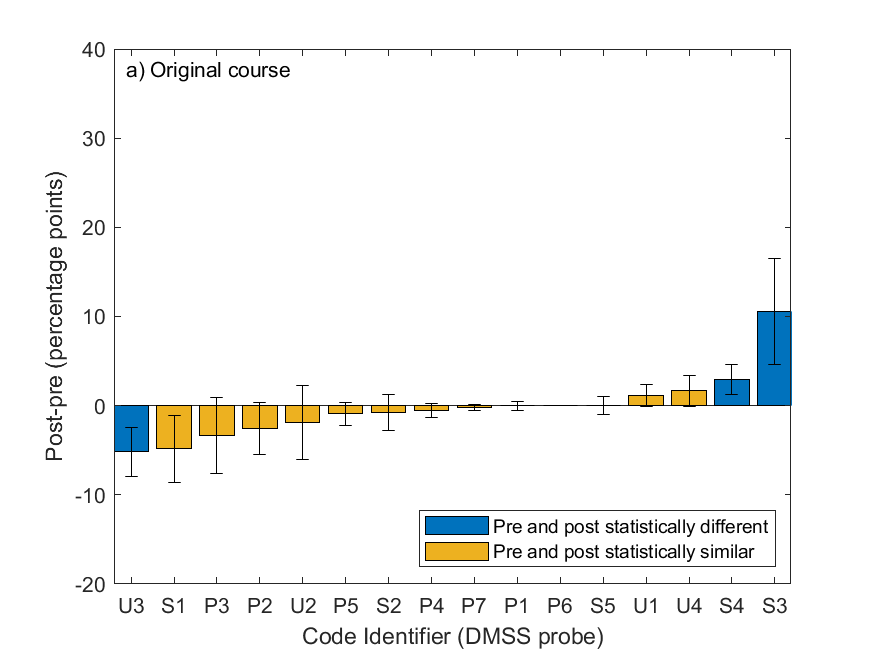}{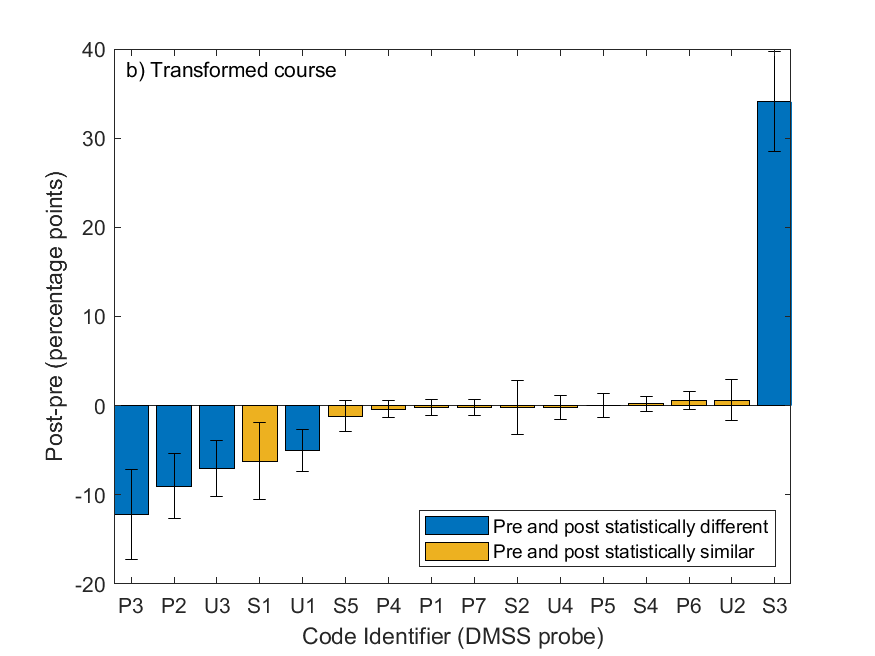}
{DMSSCodes}{Pre-post differences in code counts for the DMSS probe, for the original course (a) and the transformed course (b). Blue bars are statistically significant differences using Fisher's Exact test at the 95\% confidence interval, adjusted using the Holm-Bonferroni method. Organge bars are are not significant by that same test. Error bars are the binomial proportion confidence interval at the 95\% confidence level.}
Lastly, a comparison of DMSS codes appears in Fig. \ref{DMSSCodes}.
In both semesters, the most prominent pre-to-post increase was S3, with an especially large increase in the transformed course.
S3 represents the most complete way to compare results under the set paradigm, by looking for overlap between the means and spreads of the two data sets.

The S4 code, choosing the best multiple choice answer but leaving the explanation field blank, also increased significantly in the original course.
There could be many reasons for that difference between the two semesters, including time limitations stemming from the different settings in which the survey was administered.
The difference in S4 responses yields little insight into student learning or the transformation.
Similarly, the U3 code, which decreased in both semesters, represents miscellaneous reasoning, and yields little insight without further qualitative interpretation of these responses.
Likewise the U1 code, which decreased in the transformed course, represents non-statistical reasoning and itself offers little insight.
That is not to say that these responses, which likely contain sophisticated reasoning about systematic effects and other experimental considerations, are not worthy of study.
They are simply outside the scope of the point and set paradigms and the focus of the PMQ.

Two remaining codes decreased in prevalence during the transformed course: P2 and P3.
Both consider only the mean of the two data sets when comparing them, P2 concluding that the means must match for the results to agree, and P3 concluding that the means are close enough to agree.
Both responses lack reasoning around distributions or statistical uncertainty.
A decrease in their prevalence after the transformation, given the absence of such a decrease before the transformation, is an indication that the transformation 
was effective.

\section{Discussion and conclusions} \label{DiscussionSection}
In this section, we synthesize the results presented above through the lens of the course transformation's learning goal around measurement uncertainty, which was, ``Students should demonstrate a set-like reasoning when evaluating measurements.''
We apply the lens of that learning goal to both the original and the transformed course.
We consider the original course, in addition to the transformed course, through that lens for two reasons: first, as a baseline of comparison for the transformed course, and second as an example of an entirely traditional physics lab course that nonetheless achieved measurable and desirable learning outcomes.

\subsection{(Q1) Effectiveness of course overall}
Regarding (Q1), \textit{Did students respond to the PMQ in ways more aligned with the set paradigm after taking the introductory lab course, compared to when they began the course?}, on each of the levels of analysis presented here, both the original and the transformed course met the learning goal to some extent. 
At the most simplified level of student-paradigms, our analysis shows increases in set-like reasoning and decreases in mixed reasoning in both semesters, which is aligned with the learning goal.
At a finer level of detail looking at paradigms probe-by-probe, each of the four PMQ probes that we analyzed showed, in both semesters, pre-to-post increases in set-like reasoning and pre-to-post decreases in point-like reasoning.
In all but two cases, these increases were statistically significant.
Finally, in the most fine-grained interpretation of the results, looking at individual codes beyond their paradigms, there were pre-to-post changes in each of the four probes that aligned with the learning goal to some extent.
In particular, significant increases in RD-S4 and DMSS-S3, and significant decreases in RD-P1 and RD-P2, were observed in both semesters, and unambiguously align with the learning goal.

\subsection{(Q2) Effectiveness of transformation}
Regarding (Q2), \textit{Did student responses to the PMQ after the transformation show greater change towards the set paragigm than resposnes before the transformation?}, there were several indications that the transformed course achieved the learning goal to a greater extent than the original course. 
There were also some indications that it did not, suggesting directions for future improvement.
However, all of these indications lie at finer levels of analysis than that of student-paradigms, in which the semesters before and after transformation appear similar in all respects.

At the level of paradigms for each probe, the RD probe showed a striking pre-to-post increase in S codes after the transformation, with an effect size of h=0.45, as compared to the corresponding increase before the transformation with an effect size of h=0.21.
This difference suggests that the course transformation was especially successful in the context of evaluating choices in data collection.

On the other hand, in the SMDS probe, we observed significant pre-to-post shifts in the original course, but not in the transformed course.
This suggests that there is more for students to learn in the transformed course around data comparison, particularly in the context presented by this probe: when the two data sets being compared have identical means but different spreads.
In such a situation, deciding whether the results agree is very simple, and once can entirely ignore the spreads of the two distributions.
Accordingly, the most common SMDS point-like code across both semesters was SMDS-P1 (18\% of all responses in the data set), which represents this simple approach.

However, there is more to consider when deciding not whether the results agree, but which result is better overall. 
The SMDS probe asks respondents to do this latter task.
With this broader task in mind, the fact that one result has a smaller spread becomes relevant, as represented by the SMDS-S2 code, the most common code of any paradigm assigned to SMDS responses (53\% of all responses in the data set).
Nonetheless, it is possible that the probe does not prompt set-like reasoning as directly as other probes, as the same means encourage students to stop there without considering the data set at a deeper level. 
For the transformed course to improve further, results from the SMDS probe suggest that students could be better supported in using set-like thinking all the time, not just when the situation lends itself to it.
Perhaps including more focused or nuanced discussions around what makes a data set better or worse would result in more favorable SMDS responses, and more importantly, further improve physics laboratory instruction.

Regarding the other two probes, less can be drawn from the paradigm-level results.
While pre-to-post shifts in the UR probe were statistically significant only in the transformed course, and in directions aligned with the learning goal, the overwhelming prevalence of S responses in all cases makes this result have little practical significance.
The effect sizes of the DMSS probe are also more favorable in the transformed course, but because this difference is due to differences in pre-test proportions rather than post-test proportions, we hesitate to interpret it further.

Finally, analyzing pre-to-post differences in each individual code yields further insight into the success of the transformation.
In the RD probe, an absence of an increase in RD-U1 after the transformation indicates not only that 
more students communicated in alignment with the learning goal (as established earlier in this section), but also that the transformation allowed them to do so with greater sophistication.
In the UR probe, a consolidation of responses into UR-S4 in the transformed course, rather than UR-S1 in the original course, is additional evidence of more sophisticated reasoning, this time regarding the idea that every number has an uncertainty.
However, decreases in UR-S3 might suggest that the transformed course also de-emphasized a more sophisticated conceptual understanding of the role of means, which could be a focus of further improvement.
In the SMDS probe, there were no clear messages from analyzing pre-post differences code-by-code, underscoring the inherent consistency of responses to this probe.
While the changes in the original course seem at first to align with the learning goal, further consideration of the reasoning they represent complicates this picture (as discussed above).
More qualitative study is needed to better understand how students are interpreting, reasoning, and responding to the SMDS probe.

In the DMSS probe, the observed decrease in DMSS-P2 and DMSS-P3 in the transformed course would suggest that the transformation was successful at encouraging set-like reasoning around data comparison.
However, the irregularities around the DMSS probe, as discussed above and in the next section, cast doubt on the full implications of this finding.

\subsection{Successes and Limitations}
We start by noting a success regarding research methodology, noting that qualitatively different insight emerged as we proceeded to each deeper level of analysis detail.
Indications that the transformation was successful at meeting the learning goal around measurement uncertainty emerged only when considering responses probe-by-probe, and evidence about the depth of that learning emerged only when investigating responses code-by-code.
More generally, these observations are merely a reminder that there is far more to learning around measurement uncertainty than is captured by the point and set paradigms.
However, this bigger picture also comes with a limitation: here, we can study only the reasoning prompted by the probes of the PMQ, which is a smaller scope than all that is important to student learning around measurement uncertainty.

Furthermore, studying responses to surveys come with limitations in general, as the administration of surveys precludes the ability to ask follow-up questions.
When applying our coding scheme in this work, we had only the written survey responses to interpret, complete with any ambiguities in what the student was actually thinking.
These ambiguities often forced us to assign U codes to such responses.
For those students, more interactive methods (such as interviews) would allow us to better distinguish their ability to communicate reasoning from the nature of their reasoning itself.

We also note that the responses we analyze came from a single institution, CU, which is a large, research-focused, highly resourced, primarily white institution of a type that is overrepresented in literature \cite{Kanim2017}.
A broader range of students and institutions is needed to determine whether these findings hold beyond CU.
Additionally, we studied student reasoning at the introductory undergraduate level, and we expect these findings to apply only to students at similar academic levels.

While we did not directly compare pre and post responses student-by-student in this analysis, we still included in our data set only matched responses, that is, only those students who completed both pre- and post-test. 
We did this because we focused our analysis on measuring the effectiveness of our course, rather than making any direct comparisons to other courses or institutions.
This framing is distinct from the motivations outlined in \cite{Nissen2019}, which calls for using statistical methods to model student responses and predict the missing responses from students with unmatched data.
However, our comparisons between the original and the transformed course, given the similarity in student demographics between the two semesters, and that they come from the same institutional context and instructor, are less affected by the bias identified in that reference. 
Furthermore, it would 
require a larger data set than currently available to apply the techniques described in ref. \cite{Nissen2019} to the nominal data of paradigm or coding designations.
Nonetheless, and unavoidably, there could be some bias in the changes we observe based on exogenous factors that affect both a student's reasoning around measurement uncertainty and their likelihood to complete both the pre- and post-tests.

Lastly, the differences in timing of the pre-test between the original and transformed courses casts some doubt if the pre-test in the transformed course is a valid baseline to which to compare the post-test of that semester.
With 64\% of students completing the pretest after the first lecture in the transformed semester, this could potentially bias, but not eliminate, any measured learning outcomes from that first lecture.
However, considering the focus on measurement uncertainty throughout the course, the first lecture is a very small portion of all of the learning opportunities that the students experienced throughout the course.
Furthermore, when answering (Q2) by comparing the transformed course to the original one, additional instructional opportunities before the pre-test in the transformed course would cause measured pre-to-post changes to have a smaller effect size than otherwise, assuming the instruction has an overall effect aligned with the learning goal.
Results from (Q1) suggest that instruction does indeed shift students toward the set paradigm overall, thus, the difference in timing would result in a decrease in apparent shifts towards set-like reasoning in the transformed course.
Given that we observe evidence for the opposite effect, the timing difference is less of a concern.
However, irregularities in the transformed course pre-test results, specifically in the DMSS probe, remain a mystery, and require further investigation before results from that probe can be taken at face value.

\section{Summary} \label{ConclusionsSection}
Here, we used the PMQ to measure the effectiveness of the introductory lab course at CU, and a recent transformation of that course.
We aimed to answer two research questions: (Q1), \textit{Did students respond to the PMQ in ways more aligned with the set paradigm after taking the introductory lab course, compared to when they began the course?}, and (Q2), \textit{Did student responses to the PMQ after the transformation show greater change towards the set paradigm than responses before the transformation?}
With regards to (Q1), we see strong evidence of PMQ responses that are more aligned with the set paradigm in the post-tests from both semesters, compared to the corresponding pre-tests, and we see this evidence at all levels of analysis, from the coarsest to the finest grain sizes.
With regards to (Q2), we see evidence that PMQ responses in the transformed course shifted pre-to-post towards more prevalent set-like reasoning compared to those from the original course.
Furthermore, we also see some evidence that the responses from the transformed course tend to shift towards more sophisticated reasoning than in the original course.
We also identified specific aspects of a sophisticated understanding of measurement uncertainty that were less apparent in responses from the transformed course than from the original course, suggesting areas for further improvement.
These findings add to the growing body of evidence that physics lab courses, even traditional ones, have value by creating opportunities for students to learn important aspects of conceptual physics and to develop expert physics practices.

\acknowledgements{
We acknowledge Daniel Bolton, Colin West, Skip Woody, Michael Schefferstein, Adam Ellzey, Michael Dubson, and Jason Bossert for their work in developing the transformed course. 
We also acknowledge Dimitri Dounas-Frazer and Laura R\'ios for their input during the course transformation, and the many student testers for their contributions to improving the labs. 
We acknowledge Dimitri Dounas-Frazer for creating the online version of the PMQ, and for Saalih Allie for his perspective and advice regarding the PMQ. 
This work is supported by the NSF under grant nos. PHYS-1734006 and DMR-1548924.
It is further supported by the office of the Associate Dean for Education of the College of Engineering and Applied Science, and the College of Arts and Sciences, at the University of Colorado Boulder.
}

\bibliography{PMQPollardBib}

\begin{thebibliography}{58}%
\makeatletter
\providecommand \@ifxundefined [1]{%
 \@ifx{#1\undefined}
}%
\providecommand \@ifnum [1]{%
 \ifnum #1\expandafter \@firstoftwo
 \else \expandafter \@secondoftwo
 \fi
}%
\providecommand \@ifx [1]{%
 \ifx #1\expandafter \@firstoftwo
 \else \expandafter \@secondoftwo
 \fi
}%
\providecommand \natexlab [1]{#1}%
\providecommand \enquote  [1]{``#1''}%
\providecommand \bibnamefont  [1]{#1}%
\providecommand \bibfnamefont [1]{#1}%
\providecommand \citenamefont [1]{#1}%
\providecommand \href@noop [0]{\@secondoftwo}%
\providecommand \href [0]{\begingroup \@sanitize@url \@href}%
\providecommand \@href[1]{\@@startlink{#1}\@@href}%
\providecommand \@@href[1]{\endgroup#1\@@endlink}%
\providecommand \@sanitize@url [0]{\catcode `\\12\catcode `\$12\catcode
  `\&12\catcode `\#12\catcode `\^12\catcode `\_12\catcode `\%12\relax}%
\providecommand \@@startlink[1]{}%
\providecommand \@@endlink[0]{}%
\providecommand \url  [0]{\begingroup\@sanitize@url \@url }%
\providecommand \@url [1]{\endgroup\@href {#1}{\urlprefix }}%
\providecommand \urlprefix  [0]{URL }%
\providecommand \Eprint [0]{\href }%
\providecommand \doibase [0]{http://dx.doi.org/}%
\providecommand \selectlanguage [0]{\@gobble}%
\providecommand \bibinfo  [0]{\@secondoftwo}%
\providecommand \bibfield  [0]{\@secondoftwo}%
\providecommand \translation [1]{[#1]}%
\providecommand \BibitemOpen [0]{}%
\providecommand \bibitemStop [0]{}%
\providecommand \bibitemNoStop [0]{.\EOS\space}%
\providecommand \EOS [0]{\spacefactor3000\relax}%
\providecommand \BibitemShut  [1]{\csname bibitem#1\endcsname}%
\let\auto@bib@innerbib\@empty
\bibitem [{\citenamefont {{PCAST STEM Undergraduate Education Working
  Group}}(2012)}]{olson2012engage}%
  \BibitemOpen
  \bibfield  {author} {\bibinfo {author} {\bibnamefont {{PCAST STEM
  Undergraduate Education Working Group}}},\ }\href
  {http://files.eric.ed.gov/fulltext/ED541511.pdf} {\emph {\bibinfo {title}
  {{Engage to Excel: Producing One Million Additional College Graduates with
  Degrees in STEM}}}}\ (\bibinfo  {publisher} {Executive Office of the
  President},\ \bibinfo {year} {2012})\BibitemShut {NoStop}%
\bibitem [{\citenamefont {{National Research
  Council}}(2012)}]{national2012discipline}%
  \BibitemOpen
  \bibfield  {author} {\bibinfo {author} {\bibnamefont {{National Research
  Council}}},\ }\href {\doibase 10.17226/13362} {\emph {\bibinfo {title}
  {{Discipline-based education research: Understanding and improving learning
  in undergraduate science and engineering}}}}\ (\bibinfo  {publisher} {Natl
  Acad Press},\ \bibinfo {address} {Washington, DC},\ \bibinfo {year}
  {2012})\BibitemShut {NoStop}%
\bibitem [{\citenamefont {{AAPT Committee on
  Laboratories}}(2014)}]{kozminski2014aapt}%
  \BibitemOpen
  \bibfield  {author} {\bibinfo {author} {\bibnamefont {{AAPT Committee on
  Laboratories}}},\ }\href
  {https://www.aapt.org/Resources/upload/LabGuidlinesDocument{\_}EBendorsed{\_}nov10.pdf}
  {\emph {\bibinfo {title} {{AAPT Recommendations for the Undergraduate Physics
  Laboratory Curriculum}}}}\ (\bibinfo  {publisher} {Am Assoc Phys Teach},\
  \bibinfo {year} {2014})\BibitemShut {NoStop}%
\bibitem [{\citenamefont {{Joint Task Force on Undergraduate Physics
  Programs}}(2016)}]{Heron2016}%
  \BibitemOpen
  \bibfield  {author} {\bibinfo {author} {\bibnamefont {{Joint Task Force on
  Undergraduate Physics Programs}}},\ }\href
  {http://www.compadre.org/JTUPP/docs/J-Tupp{\_}Report.pdf} {\emph {\bibinfo
  {title} {{Phys21: Preparing Physics Students for 21st Century Careers}}}}\
  (\bibinfo  {publisher} {Am Phys Soc, Am Assoc Phys Teach},\ \bibinfo
  {address} {College Park, MD},\ \bibinfo {year} {2016})\BibitemShut {NoStop}%
\bibitem [{\citenamefont {Moskovitz}\ and\ \citenamefont
  {Kellogg}(2011)}]{Moskovitz2011}%
  \BibitemOpen
  \bibfield  {author} {\bibinfo {author} {\bibfnamefont {C.}~\bibnamefont
  {Moskovitz}}\ and\ \bibinfo {author} {\bibfnamefont {D.}~\bibnamefont
  {Kellogg}},\ }\href {\doibase 10.1126/science.1200353} {\bibfield  {journal}
  {\bibinfo  {journal} {Science}\ }\textbf {\bibinfo {volume} {332}},\ \bibinfo
  {pages} {919} (\bibinfo {year} {2011})}\BibitemShut {NoStop}%
\bibitem [{\citenamefont {Zwickl}\ \emph {et~al.}(2013)\citenamefont {Zwickl},
  \citenamefont {Finkelstein},\ and\ \citenamefont
  {Lewandowski}}]{Zwickl2013b}%
  \BibitemOpen
  \bibfield  {author} {\bibinfo {author} {\bibfnamefont {B.~M.}\ \bibnamefont
  {Zwickl}}, \bibinfo {author} {\bibfnamefont {N.}~\bibnamefont {Finkelstein}},
  \ and\ \bibinfo {author} {\bibfnamefont {H.~J.}\ \bibnamefont
  {Lewandowski}},\ }\href {\doibase 10.1119/1.4768890} {\bibfield  {journal}
  {\bibinfo  {journal} {American Journal of Physics}\ }\textbf {\bibinfo
  {volume} {81}},\ \bibinfo {pages} {63} (\bibinfo {year} {2013})},\ \Eprint
  {http://arxiv.org/abs/1207.2177} {arXiv:1207.2177} \BibitemShut {NoStop}%
\bibitem [{\citenamefont {Wilcox}\ and\ \citenamefont
  {Lewandowski}(2018)}]{Wilcox2018}%
  \BibitemOpen
  \bibfield  {author} {\bibinfo {author} {\bibfnamefont {B.~R.}\ \bibnamefont
  {Wilcox}}\ and\ \bibinfo {author} {\bibfnamefont {H.~J.}\ \bibnamefont
  {Lewandowski}},\ }\href {\doibase 10.1119/1.5009241} {\bibfield  {journal}
  {\bibinfo  {journal} {Am. J. Phys.}\ }\textbf {\bibinfo {volume} {86}},\
  \bibinfo {pages} {212} (\bibinfo {year} {2018})}\BibitemShut {NoStop}%
\bibitem [{\citenamefont {Smith}\ \emph
  {et~al.}(2020{\natexlab{a}})\citenamefont {Smith}, \citenamefont
  {Chodkowski},\ and\ \citenamefont {Holmes}}]{Smith2020}%
  \BibitemOpen
  \bibfield  {author} {\bibinfo {author} {\bibfnamefont {E.~M.}\ \bibnamefont
  {Smith}}, \bibinfo {author} {\bibfnamefont {N.}~\bibnamefont {Chodkowski}}, \
  and\ \bibinfo {author} {\bibfnamefont {N.~G.}\ \bibnamefont {Holmes}},\ }in\
  \href {\doibase 10.1119/perc.2019.pr.Smith_E} {\emph {\bibinfo {booktitle}
  {2019 Physics Education Research Conference Proceedings}}}\ (\bibinfo
  {publisher} {American Association of Physics Teachers},\ \bibinfo {year}
  {2020})\BibitemShut {NoStop}%
\bibitem [{\citenamefont {Smith}\ \emph
  {et~al.}(2020{\natexlab{b}})\citenamefont {Smith}, \citenamefont {Stein},
  \citenamefont {Walsh},\ and\ \citenamefont {Holmes}}]{Smith2020a}%
  \BibitemOpen
  \bibfield  {author} {\bibinfo {author} {\bibfnamefont {E.~M.}\ \bibnamefont
  {Smith}}, \bibinfo {author} {\bibfnamefont {M.~M.}\ \bibnamefont {Stein}},
  \bibinfo {author} {\bibfnamefont {C.}~\bibnamefont {Walsh}}, \ and\ \bibinfo
  {author} {\bibfnamefont {N.}~\bibnamefont {Holmes}},\ }\href {\doibase
  10.1103/PhysRevX.10.011029} {\bibfield  {journal} {\bibinfo  {journal}
  {Physical Review X}\ }\textbf {\bibinfo {volume} {10}},\ \bibinfo {pages}
  {011029} (\bibinfo {year} {2020}{\natexlab{b}})}\BibitemShut {NoStop}%
\bibitem [{\citenamefont {Pollard}\ \emph {et~al.}(2020)\citenamefont
  {Pollard}, \citenamefont {Hobbs}, \citenamefont {Dounas-Frazer},\ and\
  \citenamefont {Lewandowski}}]{Pollard2020}%
  \BibitemOpen
  \bibfield  {author} {\bibinfo {author} {\bibfnamefont {B.}~\bibnamefont
  {Pollard}}, \bibinfo {author} {\bibfnamefont {R.}~\bibnamefont {Hobbs}},
  \bibinfo {author} {\bibfnamefont {D.~R.}\ \bibnamefont {Dounas-Frazer}}, \
  and\ \bibinfo {author} {\bibfnamefont {H.~J.}\ \bibnamefont {Lewandowski}},\
  }in\ \href {\doibase 10.1119/perc.2019.pr.Pollard} {\emph {\bibinfo
  {booktitle} {2019 Physics Education Research Conference Proceedings}}}\
  (\bibinfo  {publisher} {American Association of Physics Teachers},\ \bibinfo
  {year} {2020})\BibitemShut {NoStop}%
\bibitem [{\citenamefont {Lewandowski}\ \emph {et~al.}(2020)\citenamefont
  {Lewandowski}, \citenamefont {Pollard},\ and\ \citenamefont
  {West}}]{Lewandowski2020}%
  \BibitemOpen
  \bibfield  {author} {\bibinfo {author} {\bibfnamefont {H.~J.}\ \bibnamefont
  {Lewandowski}}, \bibinfo {author} {\bibfnamefont {B.}~\bibnamefont
  {Pollard}}, \ and\ \bibinfo {author} {\bibfnamefont {C.~G.}\ \bibnamefont
  {West}},\ }in\ \href {\doibase 10.1119/perc.2019.pr.Lewandowski} {\emph
  {\bibinfo {booktitle} {2019 Physics Education Research Conference
  Proceedings}}}\ (\bibinfo  {publisher} {American Association of Physics
  Teachers},\ \bibinfo {year} {2020})\BibitemShut {NoStop}%
\bibitem [{\citenamefont {Lewandowski}\ \emph {et~al.}(2019)\citenamefont
  {Lewandowski}, \citenamefont {Bolton},\ and\ \citenamefont
  {Pollard}}]{Lewandowski2018}%
  \BibitemOpen
  \bibfield  {author} {\bibinfo {author} {\bibfnamefont {H.~J.}\ \bibnamefont
  {Lewandowski}}, \bibinfo {author} {\bibfnamefont {D.~R.}\ \bibnamefont
  {Bolton}}, \ and\ \bibinfo {author} {\bibfnamefont {B.}~\bibnamefont
  {Pollard}},\ }in\ \href {\doibase 10.1119/perc.2018.pr.Lewandowski} {\emph
  {\bibinfo {booktitle} {2018 Physics Education Research Conference
  Proceedings}}}\ (\bibinfo  {publisher} {American Association of Physics
  Teachers},\ \bibinfo {year} {2019})\BibitemShut {NoStop}%
\bibitem [{\citenamefont {Holmes}\ \emph {et~al.}(2017)\citenamefont {Holmes},
  \citenamefont {Olsen}, \citenamefont {Thomas},\ and\ \citenamefont
  {Wieman}}]{Holmes2017}%
  \BibitemOpen
  \bibfield  {author} {\bibinfo {author} {\bibfnamefont {N.}~\bibnamefont
  {Holmes}}, \bibinfo {author} {\bibfnamefont {J.}~\bibnamefont {Olsen}},
  \bibinfo {author} {\bibfnamefont {J.~L.}\ \bibnamefont {Thomas}}, \ and\
  \bibinfo {author} {\bibfnamefont {C.~E.}\ \bibnamefont {Wieman}},\ }\href
  {\doibase 10.1103/PhysRevPhysEducRes.13.010129} {\bibfield  {journal}
  {\bibinfo  {journal} {Physical Review Physics Education Research}\ }\textbf
  {\bibinfo {volume} {13}},\ \bibinfo {pages} {010129} (\bibinfo {year}
  {2017})}\BibitemShut {NoStop}%
\bibitem [{\citenamefont {Holmes}\ and\ \citenamefont
  {Wieman}(2018)}]{Holmes2018}%
  \BibitemOpen
  \bibfield  {author} {\bibinfo {author} {\bibfnamefont {N.~G.}\ \bibnamefont
  {Holmes}}\ and\ \bibinfo {author} {\bibfnamefont {C.~E.}\ \bibnamefont
  {Wieman}},\ }\href {\doibase 10.1063/PT.3.3816} {\bibfield  {journal}
  {\bibinfo  {journal} {Physics Today}\ }\textbf {\bibinfo {volume} {71}},\
  \bibinfo {pages} {38} (\bibinfo {year} {2018})}\BibitemShut {NoStop}%
\bibitem [{\citenamefont {Fox}\ \emph {et~al.}(2020)\citenamefont {Fox},
  \citenamefont {Werth}, \citenamefont {Hoehn},\ and\ \citenamefont
  {Lewandowski}}]{Fox2020}%
  \BibitemOpen
  \bibfield  {author} {\bibinfo {author} {\bibfnamefont {M.~F.~J.}\
  \bibnamefont {Fox}}, \bibinfo {author} {\bibfnamefont {A.}~\bibnamefont
  {Werth}}, \bibinfo {author} {\bibfnamefont {J.~R.}\ \bibnamefont {Hoehn}}, \
  and\ \bibinfo {author} {\bibfnamefont {H.~J.}\ \bibnamefont {Lewandowski}},\
  }\href {http://arxiv.org/abs/2007.01271} {\bibfield  {journal} {\bibinfo
  {journal} {arXiv.org}\ } (\bibinfo {year} {2020})},\ \Eprint
  {http://arxiv.org/abs/2007.01271} {arXiv:2007.01271} \BibitemShut {NoStop}%
\bibitem [{\citenamefont {Lippmann}(2003)}]{Lippmann2003}%
  \BibitemOpen
  \bibfield  {author} {\bibinfo {author} {\bibfnamefont {R.~F.}\ \bibnamefont
  {Lippmann}},\ }\emph {\bibinfo {title} {{Students' understanding of
  measurement and uncertainty in the physics laboratory : social construction,
  underlying concepts, and quantitative analysis}}},\ \href@noop {} {Ph.D.
  thesis},\ \bibinfo  {school} {University of Maryland, College Park} (\bibinfo
  {year} {2003})\BibitemShut {NoStop}%
\bibitem [{\citenamefont {Kung}(2005)}]{Kung2005a}%
  \BibitemOpen
  \bibfield  {author} {\bibinfo {author} {\bibfnamefont {R.~L.}\ \bibnamefont
  {Kung}},\ }\href {\doibase 10.1119/1.1881253} {\bibfield  {journal} {\bibinfo
   {journal} {American Journal of Physics}\ }\textbf {\bibinfo {volume} {73}},\
  \bibinfo {pages} {771} (\bibinfo {year} {2005})}\BibitemShut {NoStop}%
\bibitem [{\citenamefont {Beichner}(2007)}]{Beichner2007}%
  \BibitemOpen
  \bibfield  {author} {\bibinfo {author} {\bibfnamefont {R.}~\bibnamefont
  {Beichner}},\ }\href {https://www.compadre.org/per/items/detail.cfm?ID=4517}
  {\bibfield  {journal} {\bibinfo  {journal} {Research-Based Reform of
  University Physics}\ }\textbf {\bibinfo {volume} {1}} (\bibinfo {year}
  {2007})}\BibitemShut {NoStop}%
\bibitem [{\citenamefont {Abbot}(2003)}]{Abbott2003}%
  \BibitemOpen
  \bibfield  {author} {\bibinfo {author} {\bibfnamefont {D.~S.}\ \bibnamefont
  {Abbot}},\ }\emph {\bibinfo {title} {{Assessing student understanding of
  measurement and uncertainty}}},\ \href@noop {} {Ph.D. thesis},\ \bibinfo
  {school} {North Carolina State University} (\bibinfo {year}
  {2003})\BibitemShut {NoStop}%
\bibitem [{\citenamefont {Etkina}\ and\ \citenamefont
  {Heuvelen}(2007)}]{Etkina2007}%
  \BibitemOpen
  \bibfield  {author} {\bibinfo {author} {\bibfnamefont {E.}~\bibnamefont
  {Etkina}}\ and\ \bibinfo {author} {\bibfnamefont {A.~V.}\ \bibnamefont
  {Heuvelen}},\ }\href {https://www.compadre.org/per/items/detail.cfm?ID=4988}
  {\bibfield  {journal} {\bibinfo  {journal} {Research-based reform of
  university physics}\ }\textbf {\bibinfo {volume} {1}},\ \bibinfo {pages} {1}
  (\bibinfo {year} {2007})}\BibitemShut {NoStop}%
\bibitem [{\citenamefont {Etkina}\ \emph {et~al.}(2010)\citenamefont {Etkina},
  \citenamefont {Karelina}, \citenamefont {Ruibal-Villasenor}, \citenamefont
  {Rosengrant}, \citenamefont {Jordan},\ and\ \citenamefont
  {Hmelo-Silver}}]{Etkina2010}%
  \BibitemOpen
  \bibfield  {author} {\bibinfo {author} {\bibfnamefont {E.}~\bibnamefont
  {Etkina}}, \bibinfo {author} {\bibfnamefont {A.}~\bibnamefont {Karelina}},
  \bibinfo {author} {\bibfnamefont {M.}~\bibnamefont {Ruibal-Villasenor}},
  \bibinfo {author} {\bibfnamefont {D.}~\bibnamefont {Rosengrant}}, \bibinfo
  {author} {\bibfnamefont {R.}~\bibnamefont {Jordan}}, \ and\ \bibinfo {author}
  {\bibfnamefont {C.~E.}\ \bibnamefont {Hmelo-Silver}},\ }\href {\doibase
  10.1080/10508400903452876} {\bibfield  {journal} {\bibinfo  {journal}
  {Journal of the Learning Sciences}\ }\textbf {\bibinfo {volume} {19}},\
  \bibinfo {pages} {54} (\bibinfo {year} {2010})}\BibitemShut {NoStop}%
\bibitem [{\citenamefont {Holmes}(2014)}]{Holmes2014a}%
  \BibitemOpen
  \bibfield  {author} {\bibinfo {author} {\bibfnamefont {N.~G.}\ \bibnamefont
  {Holmes}},\ }\emph {\bibinfo {title} {{Structured quantitative inquiry labs :
  developing critical thinking in the introductory physics laboratory}}},\
  \href
  {https://www.semanticscholar.org/paper/Structured-quantitative-inquiry-labs-{\%}3A-developing-Holmes/e146356c79b2b93119ab1e527bb74cb2b41d4c66}
  {Ph.D. thesis},\ \bibinfo  {school} {The University of British Columbia}
  (\bibinfo {year} {2014})\BibitemShut {NoStop}%
\bibitem [{\citenamefont {Strubbe}\ \emph {et~al.}(2016)\citenamefont
  {Strubbe}, \citenamefont {Ives}, \citenamefont {Holmes}, \citenamefont
  {Bonn},\ and\ \citenamefont {Sumah}}]{Strubbe2016}%
  \BibitemOpen
  \bibfield  {author} {\bibinfo {author} {\bibfnamefont {L.~E.}\ \bibnamefont
  {Strubbe}}, \bibinfo {author} {\bibfnamefont {J.}~\bibnamefont {Ives}},
  \bibinfo {author} {\bibfnamefont {N.~G.}\ \bibnamefont {Holmes}}, \bibinfo
  {author} {\bibfnamefont {D.~A.}\ \bibnamefont {Bonn}}, \ and\ \bibinfo
  {author} {\bibfnamefont {N.~K.}\ \bibnamefont {Sumah}}\ }(\bibinfo
  {publisher} {American Association of Physics Teachers (AAPT)},\ \bibinfo
  {year} {2016})\ pp.\ \bibinfo {pages} {340--343}\BibitemShut {NoStop}%
\bibitem [{\citenamefont {Holmes}\ and\ \citenamefont
  {Smith}(2019)}]{Holmes2019}%
  \BibitemOpen
  \bibfield  {author} {\bibinfo {author} {\bibfnamefont {N.~G.}\ \bibnamefont
  {Holmes}}\ and\ \bibinfo {author} {\bibfnamefont {E.~M.}\ \bibnamefont
  {Smith}},\ }\href {\doibase 10.1119/1.5098916} {\bibfield  {journal}
  {\bibinfo  {journal} {The Physics Teacher}\ }\textbf {\bibinfo {volume}
  {57}},\ \bibinfo {pages} {296} (\bibinfo {year} {2019})}\BibitemShut
  {NoStop}%
\bibitem [{\citenamefont {Smith}\ \emph
  {et~al.}(2020{\natexlab{c}})\citenamefont {Smith}, \citenamefont {Stein},
  \citenamefont {Walsh},\ and\ \citenamefont {Holmes}}]{Smith2020b}%
  \BibitemOpen
  \bibfield  {author} {\bibinfo {author} {\bibfnamefont {E.~M.}\ \bibnamefont
  {Smith}}, \bibinfo {author} {\bibfnamefont {M.~M.}\ \bibnamefont {Stein}},
  \bibinfo {author} {\bibfnamefont {C.}~\bibnamefont {Walsh}}, \ and\ \bibinfo
  {author} {\bibfnamefont {N.~G.}\ \bibnamefont {Holmes}},\ }\href {\doibase
  10.1103/PhysRevX.10.011029} {\bibfield  {journal} {\bibinfo  {journal}
  {Physical Review X}\ }\textbf {\bibinfo {volume} {10}},\ \bibinfo {pages}
  {011029} (\bibinfo {year} {2020}{\natexlab{c}})}\BibitemShut {NoStop}%
\bibitem [{\citenamefont {Deardorff}(2001)}]{Deardorff2001}%
  \BibitemOpen
  \bibfield  {author} {\bibinfo {author} {\bibfnamefont {D.~L.}\ \bibnamefont
  {Deardorff}},\ }\emph {\bibinfo {title} {{Introductory physics students'
  treatment of measurement uncertainty}}},\ \href
  {http://isobudgets.com/pdf/papers/10{\_}DeardorffDissertation.pdf
  file:///Users/Kuo/Documents/Papers/Deardorff/Deardorff 2001
  Dissertation.pdf{\%}5Cnpapers2://publication/uuid/5CA0E18E-A17F-4CBD-A120-0D28654DE3DC}
  {Ph.D. thesis},\ \bibinfo  {school} {North Carolina State University}
  (\bibinfo {year} {2001})\BibitemShut {NoStop}%
\bibitem [{\citenamefont {{Lippmann Kung}}\ and\ \citenamefont
  {Linder}(2006)}]{LippmannKung2006}%
  \BibitemOpen
  \bibfield  {author} {\bibinfo {author} {\bibfnamefont {R.}~\bibnamefont
  {{Lippmann Kung}}}\ and\ \bibinfo {author} {\bibfnamefont {C.}~\bibnamefont
  {Linder}},\ }\href@noop {} {\bibfield  {journal} {\bibinfo  {journal}
  {NorDiNa}\ }\textbf {\bibinfo {volume} {4}},\ \bibinfo {pages} {40} (\bibinfo
  {year} {2006})}\BibitemShut {NoStop}%
\bibitem [{\citenamefont {Holmes}\ and\ \citenamefont
  {Bonn}(2014)}]{Holmes2014b}%
  \BibitemOpen
  \bibfield  {author} {\bibinfo {author} {\bibfnamefont {N.~G.}\ \bibnamefont
  {Holmes}}\ and\ \bibinfo {author} {\bibfnamefont {D.~A.}\ \bibnamefont
  {Bonn}},\ }in\ \href {\doibase 10.1119/perc.2013.pr.034} {\emph {\bibinfo
  {booktitle} {Physics Education Research Conference Proceedings}}}\ (\bibinfo
  {publisher} {American Association of Physics Teachers (AAPT)},\ \bibinfo
  {year} {2014})\ pp.\ \bibinfo {pages} {185--188}\BibitemShut {NoStop}%
\bibitem [{\citenamefont {Majiet}\ and\ \citenamefont
  {Allie}(2018)}]{Majiet2018}%
  \BibitemOpen
  \bibfield  {author} {\bibinfo {author} {\bibfnamefont {N.}~\bibnamefont
  {Majiet}}\ and\ \bibinfo {author} {\bibfnamefont {S.}~\bibnamefont {Allie}},\
  }in\ \href {\doibase 10.1119/perc.2018.pr.majiet} {\emph {\bibinfo
  {booktitle} {Physics Education Research Conference Proceedings}}},\ Vol.\
  \bibinfo {volume} {2018}\ (\bibinfo  {publisher} {American Association of
  Physics Teachers},\ \bibinfo {year} {2018})\BibitemShut {NoStop}%
\bibitem [{\citenamefont {Stein}\ \emph {et~al.}(2020)\citenamefont {Stein},
  \citenamefont {White}, \citenamefont {Passante},\ and\ \citenamefont
  {Holmes}}]{Stein2020}%
  \BibitemOpen
  \bibfield  {author} {\bibinfo {author} {\bibfnamefont {M.~M.}\ \bibnamefont
  {Stein}}, \bibinfo {author} {\bibfnamefont {C.}~\bibnamefont {White}},
  \bibinfo {author} {\bibfnamefont {G.}~\bibnamefont {Passante}}, \ and\
  \bibinfo {author} {\bibfnamefont {N.~G.}\ \bibnamefont {Holmes}},\ }in\ \href
  {\doibase 10.1119/perc.2019.pr.stein} {\emph {\bibinfo {booktitle} {Physics
  Education Research Conference Proceedings}}}\ (\bibinfo  {publisher}
  {American Association of Physics Teachers (AAPT)},\ \bibinfo {year}
  {2020})\BibitemShut {NoStop}%
\bibitem [{\citenamefont {Madsen}\ \emph {et~al.}(2019)\citenamefont {Madsen},
  \citenamefont {McKagan}, \citenamefont {Sayre},\ and\ \citenamefont
  {Paul}}]{Madsen2019}%
  \BibitemOpen
  \bibfield  {author} {\bibinfo {author} {\bibfnamefont {A.}~\bibnamefont
  {Madsen}}, \bibinfo {author} {\bibfnamefont {S.~B.}\ \bibnamefont {McKagan}},
  \bibinfo {author} {\bibfnamefont {E.~C.}\ \bibnamefont {Sayre}}, \ and\
  \bibinfo {author} {\bibfnamefont {C.~A.}\ \bibnamefont {Paul}},\ }\href
  {\doibase 10.1119/1.5094139} {\bibfield  {journal} {\bibinfo  {journal}
  {American Journal of Physics}\ }\textbf {\bibinfo {volume} {87}},\ \bibinfo
  {pages} {350} (\bibinfo {year} {2019})}\BibitemShut {NoStop}%
\bibitem [{\citenamefont {Pollard}\ \emph {et~al.}(2018)\citenamefont
  {Pollard}, \citenamefont {Hobbs}, \citenamefont {Stanley}, \citenamefont
  {Dounas-Frazer},\ and\ \citenamefont {Lewandowski}}]{Pollard2017}%
  \BibitemOpen
  \bibfield  {author} {\bibinfo {author} {\bibfnamefont {B.}~\bibnamefont
  {Pollard}}, \bibinfo {author} {\bibfnamefont {R.}~\bibnamefont {Hobbs}},
  \bibinfo {author} {\bibfnamefont {J.~T.}\ \bibnamefont {Stanley}}, \bibinfo
  {author} {\bibfnamefont {D.~R.}\ \bibnamefont {Dounas-Frazer}}, \ and\
  \bibinfo {author} {\bibfnamefont {H.~J.}\ \bibnamefont {Lewandowski}},\ }in\
  \href {\doibase 10.1119/perc.2017.pr.073} {\emph {\bibinfo {booktitle} {2017
  Physics Education Research Conference Proceedings}}}\ (\bibinfo  {publisher}
  {American Association of Physics Teachers},\ \bibinfo {year} {2018})\ pp.\
  \bibinfo {pages} {312--315}\BibitemShut {NoStop}%
\bibitem [{\citenamefont {Day}\ and\ \citenamefont {Bonn}(2011)}]{Day2011a}%
  \BibitemOpen
  \bibfield  {author} {\bibinfo {author} {\bibfnamefont {J.}~\bibnamefont
  {Day}}\ and\ \bibinfo {author} {\bibfnamefont {D.}~\bibnamefont {Bonn}},\
  }\href {\doibase 10.1103/PhysRevSTPER.7.010114} {\bibfield  {journal}
  {\bibinfo  {journal} {Physical Review Physics Education Research}\ }\textbf
  {\bibinfo {volume} {7}},\ \bibinfo {pages} {010114} (\bibinfo {year}
  {2011})}\BibitemShut {NoStop}%
\bibitem [{\citenamefont {Holmes}\ \emph {et~al.}(2014)\citenamefont {Holmes},
  \citenamefont {Day}, \citenamefont {Park}, \citenamefont {Bonn},\ and\
  \citenamefont {Roll}}]{Holmes2014}%
  \BibitemOpen
  \bibfield  {author} {\bibinfo {author} {\bibfnamefont {N.~G.}\ \bibnamefont
  {Holmes}}, \bibinfo {author} {\bibfnamefont {J.}~\bibnamefont {Day}},
  \bibinfo {author} {\bibfnamefont {A.~H.~K.}\ \bibnamefont {Park}}, \bibinfo
  {author} {\bibfnamefont {D.~A.}\ \bibnamefont {Bonn}}, \ and\ \bibinfo
  {author} {\bibfnamefont {I.}~\bibnamefont {Roll}},\ }\href {\doibase
  10.1007/s11251-013-9300-7} {\bibfield  {journal} {\bibinfo  {journal} {Instr
  Sci}\ }\textbf {\bibinfo {volume} {42}},\ \bibinfo {pages} {523} (\bibinfo
  {year} {2014})}\BibitemShut {NoStop}%
\bibitem [{\citenamefont {Day}\ \emph {et~al.}(2016)\citenamefont {Day},
  \citenamefont {Stang}, \citenamefont {Holmes}, \citenamefont {Kumar},\ and\
  \citenamefont {Bonn}}]{Day2016}%
  \BibitemOpen
  \bibfield  {author} {\bibinfo {author} {\bibfnamefont {J.}~\bibnamefont
  {Day}}, \bibinfo {author} {\bibfnamefont {J.~B.}\ \bibnamefont {Stang}},
  \bibinfo {author} {\bibfnamefont {N.~G.}\ \bibnamefont {Holmes}}, \bibinfo
  {author} {\bibfnamefont {D.}~\bibnamefont {Kumar}}, \ and\ \bibinfo {author}
  {\bibfnamefont {D.~A.}\ \bibnamefont {Bonn}},\ }\href {\doibase
  10.1103/PhysRevPhysEducRes.12.020104} {\bibfield  {journal} {\bibinfo
  {journal} {Phys Rev Spec Top-PH}\ }\textbf {\bibinfo {volume} {12}},\
  \bibinfo {pages} {020104} (\bibinfo {year} {2016})}\BibitemShut {NoStop}%
\bibitem [{\citenamefont {Eshach}\ and\ \citenamefont
  {Kukliansky}(2016)}]{Eshach2016}%
  \BibitemOpen
  \bibfield  {author} {\bibinfo {author} {\bibfnamefont {H.}~\bibnamefont
  {Eshach}}\ and\ \bibinfo {author} {\bibfnamefont {I.}~\bibnamefont
  {Kukliansky}},\ }\href {\doibase 10.1139/cjp-2016-0308} {\bibfield  {journal}
  {\bibinfo  {journal} {Can J Phys}\ }\textbf {\bibinfo {volume} {94}},\
  \bibinfo {pages} {1205} (\bibinfo {year} {2016})}\BibitemShut {NoStop}%
\bibitem [{\citenamefont {Walsh}\ \emph {et~al.}(2019)\citenamefont {Walsh},
  \citenamefont {Quinn}, \citenamefont {Wieman},\ and\ \citenamefont
  {Holmes}}]{Walsh2019}%
  \BibitemOpen
  \bibfield  {author} {\bibinfo {author} {\bibfnamefont {C.}~\bibnamefont
  {Walsh}}, \bibinfo {author} {\bibfnamefont {K.~N.}\ \bibnamefont {Quinn}},
  \bibinfo {author} {\bibfnamefont {C.}~\bibnamefont {Wieman}}, \ and\ \bibinfo
  {author} {\bibfnamefont {N.}~\bibnamefont {Holmes}},\ }\href {\doibase
  10.1103/PhysRevPhysEducRes.15.010135} {\bibfield  {journal} {\bibinfo
  {journal} {Physical Review Physics Education Research}\ }\textbf {\bibinfo
  {volume} {15}},\ \bibinfo {pages} {010135} (\bibinfo {year}
  {2019})}\BibitemShut {NoStop}%
\bibitem [{\citenamefont {Campbell}\ \emph {et~al.}(2005)\citenamefont
  {Campbell}, \citenamefont {Lubben}, \citenamefont {Buffler},\ and\
  \citenamefont {Allie}}]{Campbell2005}%
  \BibitemOpen
  \bibfield  {author} {\bibinfo {author} {\bibfnamefont {B.}~\bibnamefont
  {Campbell}}, \bibinfo {author} {\bibfnamefont {F.}~\bibnamefont {Lubben}},
  \bibinfo {author} {\bibfnamefont {A.}~\bibnamefont {Buffler}}, \ and\
  \bibinfo {author} {\bibfnamefont {S.}~\bibnamefont {Allie}},\ }\href
  {http://www.phy.uct.ac.za/sites/default/files/image{\_}tool/images/281/people/buffler/physics{\_}education/Monograph
  2005.pdf} {\bibfield  {journal} {\bibinfo  {journal} {AJRMSTE}\ ,\ \bibinfo
  {pages} {1}} (\bibinfo {year} {2005})}\BibitemShut {NoStop}%
\bibitem [{\citenamefont {Pollard}\ and\ \citenamefont
  {Lewandowski}(2019)}]{Pollard2018}%
  \BibitemOpen
  \bibfield  {author} {\bibinfo {author} {\bibfnamefont {B.}~\bibnamefont
  {Pollard}}\ and\ \bibinfo {author} {\bibfnamefont {H.~J.}\ \bibnamefont
  {Lewandowski}},\ }in\ \href {\doibase 10.1119/perc.2018.pr.Pollard} {\emph
  {\bibinfo {booktitle} {2018 Physics Education Research Conference
  Proceedings}}}\ (\bibinfo  {publisher} {American Association of Physics
  Teachers},\ \bibinfo {year} {2019})\BibitemShut {NoStop}%
\bibitem [{\citenamefont {Lewandowski}\ \emph {et~al.}(2018)\citenamefont
  {Lewandowski}, \citenamefont {Hobbs}, \citenamefont {Stanley}, \citenamefont
  {Dounas-Frazer},\ and\ \citenamefont {Pollard}}]{Lewandowski2017}%
  \BibitemOpen
  \bibfield  {author} {\bibinfo {author} {\bibfnamefont {H.~J.}\ \bibnamefont
  {Lewandowski}}, \bibinfo {author} {\bibfnamefont {R.}~\bibnamefont {Hobbs}},
  \bibinfo {author} {\bibfnamefont {J.~T.}\ \bibnamefont {Stanley}}, \bibinfo
  {author} {\bibfnamefont {D.~R.}\ \bibnamefont {Dounas-Frazer}}, \ and\
  \bibinfo {author} {\bibfnamefont {B.}~\bibnamefont {Pollard}},\ }in\ \href
  {\doibase 10.1119/perc.2017.pr.056} {\emph {\bibinfo {booktitle} {2017
  Physics Education Research Conference Proceedings}}}\ (\bibinfo  {publisher}
  {American Association of Physics Teachers},\ \bibinfo {year} {2018})\ pp.\
  \bibinfo {pages} {244--247}\BibitemShut {NoStop}%
\bibitem [{\citenamefont {Lubben}\ and\ \citenamefont
  {Millar}(1996)}]{Lubben1996}%
  \BibitemOpen
  \bibfield  {author} {\bibinfo {author} {\bibfnamefont {F.}~\bibnamefont
  {Lubben}}\ and\ \bibinfo {author} {\bibfnamefont {R.}~\bibnamefont
  {Millar}},\ }\href {\doibase 10.1080/0950069960180807} {\bibfield  {journal}
  {\bibinfo  {journal} {International Journal of Science Education}\ }\textbf
  {\bibinfo {volume} {18}},\ \bibinfo {pages} {955} (\bibinfo {year}
  {1996})}\BibitemShut {NoStop}%
\bibitem [{\citenamefont {Volkwyn}(2005)}]{Volkwyn2005}%
  \BibitemOpen
  \bibfield  {author} {\bibinfo {author} {\bibfnamefont {T.~S.}\ \bibnamefont
  {Volkwyn}},\ }\emph {\bibinfo {title} {{First year students' understanding of
  measurement in physics laboratory work}}},\ \href@noop {} {Ph.D. thesis}
  (\bibinfo {year} {2005})\BibitemShut {NoStop}%
\bibitem [{\citenamefont {Millar}\ \emph {et~al.}(1994)\citenamefont {Millar},
  \citenamefont {Lubben}, \citenamefont {Gott},\ and\ \citenamefont
  {Duggan}}]{Millar1994}%
  \BibitemOpen
  \bibfield  {author} {\bibinfo {author} {\bibfnamefont {R.}~\bibnamefont
  {Millar}}, \bibinfo {author} {\bibfnamefont {F.}~\bibnamefont {Lubben}},
  \bibinfo {author} {\bibfnamefont {R.}~\bibnamefont {Gott}}, \ and\ \bibinfo
  {author} {\bibfnamefont {S.}~\bibnamefont {Duggan}},\ }\href {\doibase
  10.1080/0267152940090205} {\bibfield  {journal} {\bibinfo  {journal}
  {Research Papers in Education}\ }\textbf {\bibinfo {volume} {9}},\ \bibinfo
  {pages} {207} (\bibinfo {year} {1994})}\BibitemShut {NoStop}%
\bibitem [{\citenamefont {Millar}\ \emph {et~al.}(1996)\citenamefont {Millar},
  \citenamefont {Gott}, \citenamefont {Lubben},\ and\ \citenamefont
  {Duggan}}]{Millar1996}%
  \BibitemOpen
  \bibfield  {author} {\bibinfo {author} {\bibfnamefont {R.}~\bibnamefont
  {Millar}}, \bibinfo {author} {\bibfnamefont {R.}~\bibnamefont {Gott}},
  \bibinfo {author} {\bibfnamefont {F.}~\bibnamefont {Lubben}}, \ and\ \bibinfo
  {author} {\bibfnamefont {S.}~\bibnamefont {Duggan}},\ }\enquote {\bibinfo
  {title} {{Children's performance in investigative tasks in science: A
  framework for considering progression}},}\ in\ \href@noop {} {\emph {\bibinfo
  {booktitle} {Progression in learning}}},\ \bibinfo {editor} {edited by\
  \bibinfo {editor} {\bibfnamefont {M.}~\bibnamefont {Hughes}}}\ (\bibinfo
  {publisher} {Multilingual Matters Ltd},\ \bibinfo {year} {1996})\ pp.\
  \bibinfo {pages} {82--108}\BibitemShut {NoStop}%
\bibitem [{\citenamefont {Buffler}\ \emph {et~al.}(2001)\citenamefont
  {Buffler}, \citenamefont {Allie},\ and\ \citenamefont
  {Lubben}}]{Buffler2001}%
  \BibitemOpen
  \bibfield  {author} {\bibinfo {author} {\bibfnamefont {A.}~\bibnamefont
  {Buffler}}, \bibinfo {author} {\bibfnamefont {S.}~\bibnamefont {Allie}}, \
  and\ \bibinfo {author} {\bibfnamefont {F.}~\bibnamefont {Lubben}},\ }\href
  {\doibase 10.1080/09500690110039567} {\bibfield  {journal} {\bibinfo
  {journal} {Int J Sci Educ}\ }\textbf {\bibinfo {volume} {23}},\ \bibinfo
  {pages} {37} (\bibinfo {year} {2001})}\BibitemShut {NoStop}%
\bibitem [{\citenamefont {Buffler}\ \emph {et~al.}(2003)\citenamefont
  {Buffler}, \citenamefont {Allie}, \citenamefont {Lubben},\ and\ \citenamefont
  {Campbell}}]{Buffler2003}%
  \BibitemOpen
  \bibfield  {author} {\bibinfo {author} {\bibfnamefont {A.}~\bibnamefont
  {Buffler}}, \bibinfo {author} {\bibfnamefont {S.}~\bibnamefont {Allie}},
  \bibinfo {author} {\bibfnamefont {F.}~\bibnamefont {Lubben}}, \ and\ \bibinfo
  {author} {\bibfnamefont {B.}~\bibnamefont {Campbell}},\ }\href@noop {}
  {\bibfield  {journal} {\bibinfo  {journal} {4th Conference of the European
  Science Education Research Association}\ }\textbf {\bibinfo {volume} {09}},\
  \bibinfo {pages} {19} (\bibinfo {year} {2003})}\BibitemShut {NoStop}%
\bibitem [{\citenamefont {Volkwyn}\ \emph {et~al.}(2008)\citenamefont
  {Volkwyn}, \citenamefont {Allie}, \citenamefont {Buffler},\ and\
  \citenamefont {Lubben}}]{Volkwyn2008b}%
  \BibitemOpen
  \bibfield  {author} {\bibinfo {author} {\bibfnamefont {T.~S.}\ \bibnamefont
  {Volkwyn}}, \bibinfo {author} {\bibfnamefont {S.}~\bibnamefont {Allie}},
  \bibinfo {author} {\bibfnamefont {A.}~\bibnamefont {Buffler}}, \ and\
  \bibinfo {author} {\bibfnamefont {F.}~\bibnamefont {Lubben}},\ }\href
  {\doibase 10.1103/PhysRevSTPER.4.010108} {\bibfield  {journal} {\bibinfo
  {journal} {Physical Review Special Topics - Physics Education Research}\
  }\textbf {\bibinfo {volume} {4}} (\bibinfo {year} {2008}),\
  10.1103/PhysRevSTPER.4.010108}\BibitemShut {NoStop}%
\bibitem [{\citenamefont {Qualtrics}(2005)}]{Qualtrics2005}%
  \BibitemOpen
  \bibfield  {author} {\bibinfo {author} {\bibnamefont {Qualtrics}},\ }\href
  {https://www.qualtrics.com} {\enquote {\bibinfo {title}
  {https://www.qualtrics.com},}\ } (\bibinfo {year} {2005})\BibitemShut
  {NoStop}%
\bibitem [{\citenamefont {Dounas-Frazer}\ and\ \citenamefont
  {Lewandowski}(2018)}]{Dounas-Frazer2018b}%
  \BibitemOpen
  \bibfield  {author} {\bibinfo {author} {\bibfnamefont {D.~R.}\ \bibnamefont
  {Dounas-Frazer}}\ and\ \bibinfo {author} {\bibfnamefont {H.~J.}\ \bibnamefont
  {Lewandowski}},\ }\href {\doibase 10.1088/1361-6404/AAE3CE} {\bibfield
  {journal} {\bibinfo  {journal} {European Journal of Physics}\ }\textbf
  {\bibinfo {volume} {39}},\ \bibinfo {pages} {064005} (\bibinfo {year}
  {2018})}\BibitemShut {NoStop}%
\bibitem [{\citenamefont {Fisher}(1922)}]{Fisher1922}%
  \BibitemOpen
  \bibfield  {author} {\bibinfo {author} {\bibfnamefont {R.~A.}\ \bibnamefont
  {Fisher}},\ }\href {\doibase 10.2307/2340521} {\bibfield  {journal} {\bibinfo
   {journal} {Journal of the Royal Statistical Society}\ }\textbf {\bibinfo
  {volume} {85}},\ \bibinfo {pages} {87} (\bibinfo {year} {1922})}\BibitemShut
  {NoStop}%
\bibitem [{\citenamefont {Holm}(1979)}]{Holm1979}%
  \BibitemOpen
  \bibfield  {author} {\bibinfo {author} {\bibfnamefont {S.}~\bibnamefont
  {Holm}},\ }\href {\doibase 10.2307/4615733} {\bibfield  {journal} {\bibinfo
  {journal} {Scandinavian Journal of Statistics}\ }\textbf {\bibinfo {volume}
  {6}},\ \bibinfo {pages} {65} (\bibinfo {year} {1979})}\BibitemShut {NoStop}%
\bibitem [{\citenamefont {{R Core Team}}(2019)}]{RCoreTeam2017}%
  \BibitemOpen
  \bibfield  {author} {\bibinfo {author} {\bibnamefont {{R Core Team}}},\
  }\href {https://www.r-project.org/} {\emph {\bibinfo {title} {{R: A Language
  and Environment for Statistical Computing}}}},\ \bibinfo {organization} {R
  Foundation for Statistical Computing},\ \bibinfo {address} {Vienna, Austria}
  (\bibinfo {year} {2019})\BibitemShut {NoStop}%
\bibitem [{\citenamefont {Cohen}(2013)}]{cohen2013statistical}%
  \BibitemOpen
  \bibfield  {author} {\bibinfo {author} {\bibfnamefont {J.}~\bibnamefont
  {Cohen}},\ }\href {https://books.google.com/books?id=cIJH0lR33bgC} {\emph
  {\bibinfo {title} {{Statistical Power Analysis for the Behavioral
  Sciences}}}},\ \bibinfo {edition} {2nd}\ ed.\ (\bibinfo  {publisher}
  {Routledge},\ \bibinfo {year} {2013})\ p.\ \bibinfo {pages} {567}\BibitemShut
  {NoStop}%
\bibitem [{\citenamefont {Parks}\ and\ \citenamefont
  {Schmeichel}(2012)}]{AmyNoelleParks2012}%
  \BibitemOpen
  \bibfield  {author} {\bibinfo {author} {\bibfnamefont {A.~N.}\ \bibnamefont
  {Parks}}\ and\ \bibinfo {author} {\bibfnamefont {M.}~\bibnamefont
  {Schmeichel}},\ }\href {\doibase 10.5951/jresematheduc.43.3.0238} {\bibfield
  {journal} {\bibinfo  {journal} {Journal for Research in Mathematics
  Education}\ }\textbf {\bibinfo {volume} {43}},\ \bibinfo {pages} {238}
  (\bibinfo {year} {2012})}\BibitemShut {NoStop}%
\bibitem [{\citenamefont {Kanim}\ and\ \citenamefont {Cid}(2020)}]{Kanim2020}%
  \BibitemOpen
  \bibfield  {author} {\bibinfo {author} {\bibfnamefont {S.}~\bibnamefont
  {Kanim}}\ and\ \bibinfo {author} {\bibfnamefont {X.}~\bibnamefont {Cid}},\
  }\href {\doibase 10.1103/PhysRevPhysEducRes.16.020106} {\bibfield  {journal}
  {\bibinfo  {journal} {Physical Review Physics Education Research}\ }\textbf
  {\bibinfo {volume} {16}},\ \bibinfo {pages} {020106} (\bibinfo {year}
  {2020})}\BibitemShut {NoStop}%
\bibitem [{\citenamefont {Wilcox}\ and\ \citenamefont
  {Lewandowski}(2016)}]{Wilcox2016a}%
  \BibitemOpen
  \bibfield  {author} {\bibinfo {author} {\bibfnamefont {B.~R.}\ \bibnamefont
  {Wilcox}}\ and\ \bibinfo {author} {\bibfnamefont {H.}~\bibnamefont
  {Lewandowski}},\ }\href {\doibase 10.1103/PhysRevPhysEducRes.12.010123}
  {\bibfield  {journal} {\bibinfo  {journal} {Physical Review Physics Education
  Research}\ }\textbf {\bibinfo {volume} {12}},\ \bibinfo {pages} {010123}
  (\bibinfo {year} {2016})}\BibitemShut {NoStop}%
\bibitem [{\citenamefont {Kanim}\ and\ \citenamefont {Cid}(2017)}]{Kanim2017}%
  \BibitemOpen
  \bibfield  {author} {\bibinfo {author} {\bibfnamefont {S.}~\bibnamefont
  {Kanim}}\ and\ \bibinfo {author} {\bibfnamefont {X.~C.}\ \bibnamefont
  {Cid}},\ }\href {http://arxiv.org/abs/1710.02598} {\bibfield  {journal}
  {\bibinfo  {journal} {arXiv.org}\ }\textbf {\bibinfo {volume} {1710.02598}}
  (\bibinfo {year} {2017})},\ \Eprint {http://arxiv.org/abs/1710.02598}
  {arXiv:1710.02598} \BibitemShut {NoStop}%
\bibitem [{\citenamefont {Nissen}\ \emph {et~al.}(2019)\citenamefont {Nissen},
  \citenamefont {Donatello},\ and\ \citenamefont {{Van Dusen}}}]{Nissen2019}%
  \BibitemOpen
  \bibfield  {author} {\bibinfo {author} {\bibfnamefont {J.}~\bibnamefont
  {Nissen}}, \bibinfo {author} {\bibfnamefont {R.}~\bibnamefont {Donatello}}, \
  and\ \bibinfo {author} {\bibfnamefont {B.}~\bibnamefont {{Van Dusen}}},\
  }\href {\doibase 10.1103/PhysRevPhysEducRes.15.020106} {\bibfield  {journal}
  {\bibinfo  {journal} {Physical Review Physics Education Research}\ }\textbf
  {\bibinfo {volume} {15}},\ \bibinfo {pages} {020106} (\bibinfo {year}
  {2019})},\ \Eprint {http://arxiv.org/abs/1809.00035} {arXiv:1809.00035}
  \BibitemShut {NoStop}%
\end{thebibliography}%

\appendix
\pagebreak
\section{PMQ Probes} \label{AppProbes}
\nopagebreak
\insertFigureFullPage{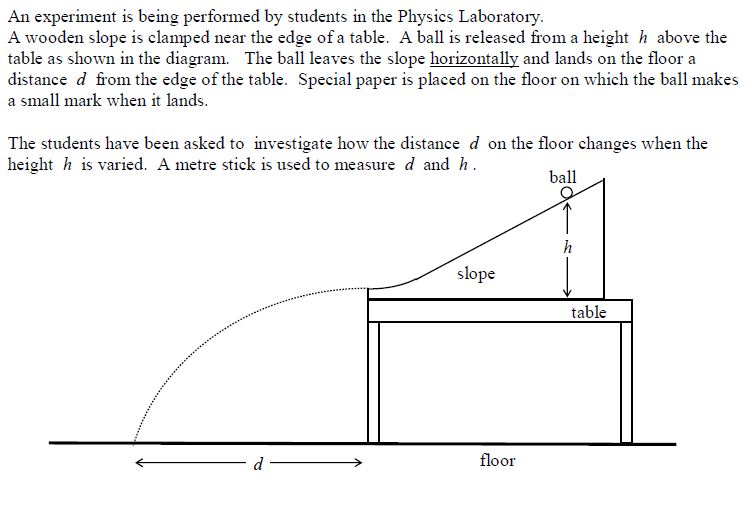}{ApdxContext}{Contextual information for the PMQ. This information precedes the probes themselves. Reproduced from ref. \cite{Volkwyn2008b}.}
\insertFigureFullPage{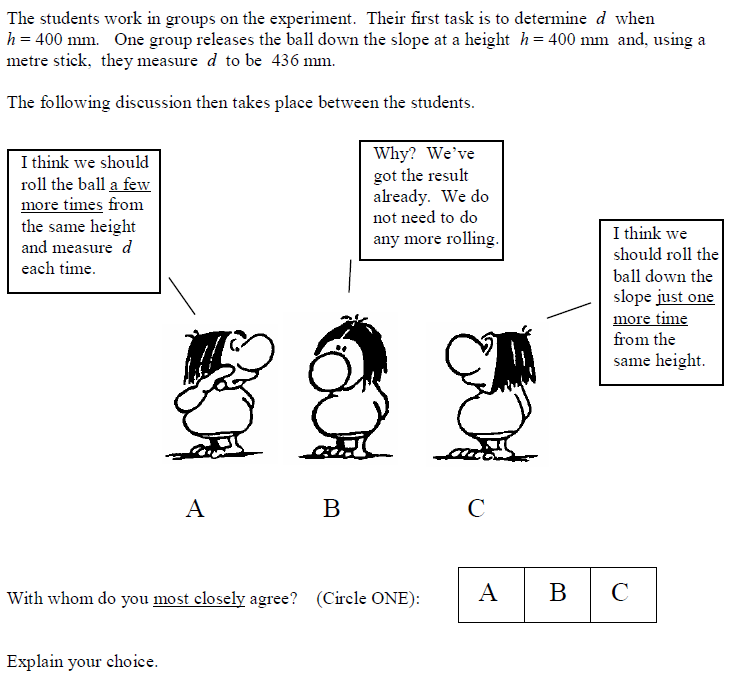}{ApdxRD}{The RD probe of the PMQ. Reproduced from ref. \cite{Volkwyn2008b}.}
\insertFigureFullPage{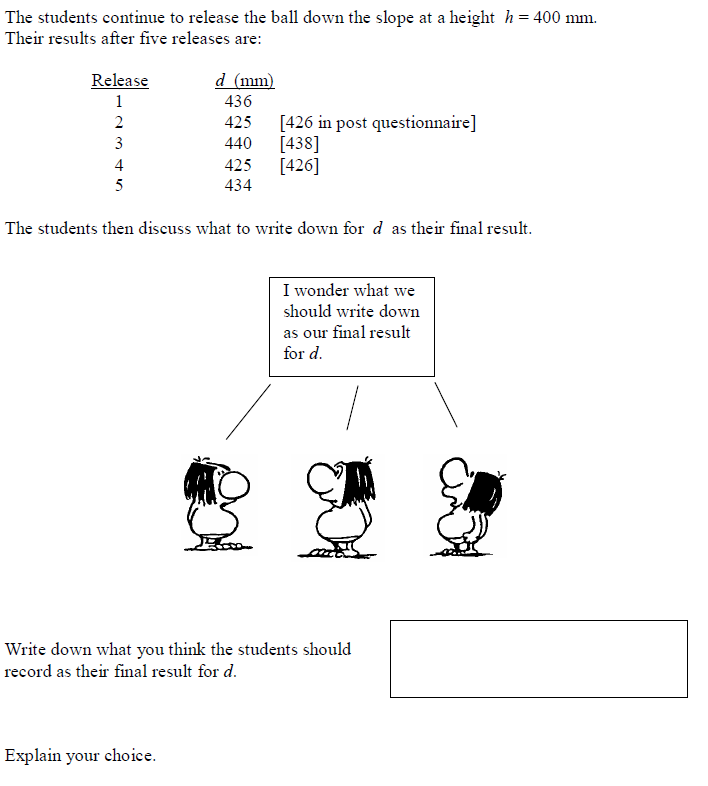}{ApdxUR}{The UR probe of the PMQ. Reproduced from ref. \cite{Volkwyn2008b}.}
\insertFigureFullPage{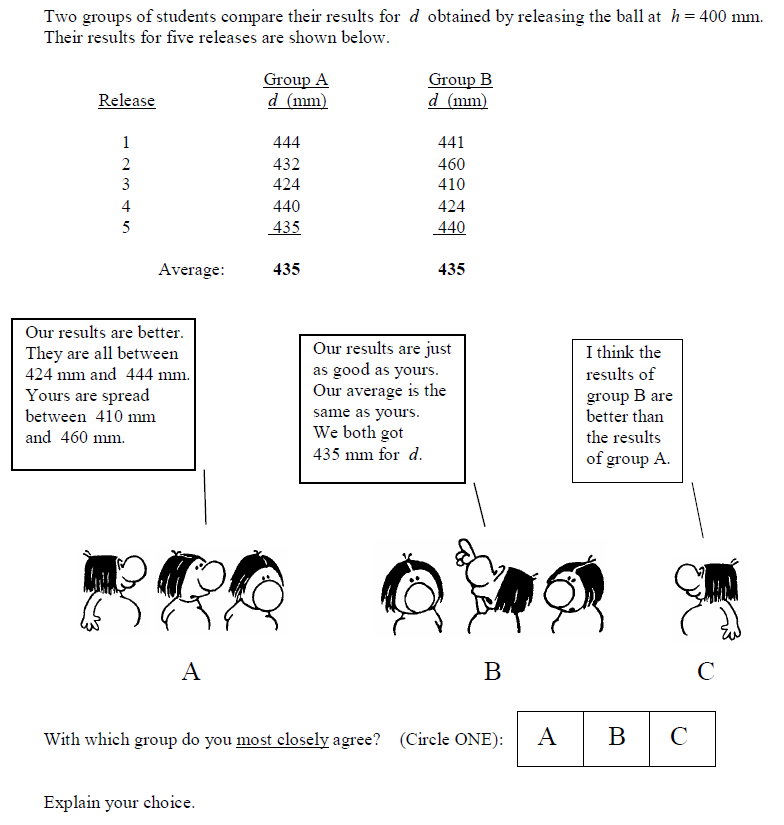}{ApdxSMDS}{The SMDS probe of the PMQ. Reproduced from ref. \cite{Volkwyn2008b}.}
\insertFigureFullPage{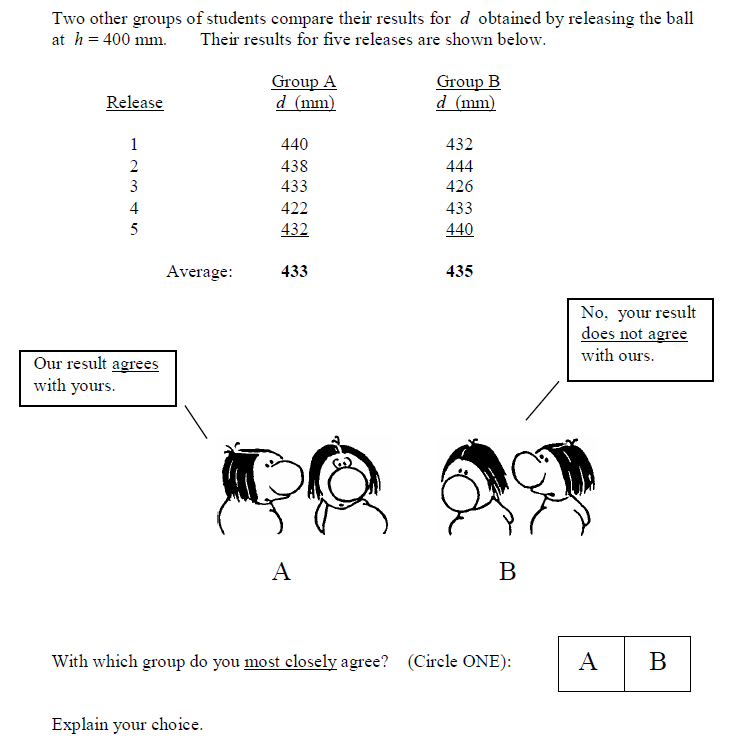}{ApdxDMSS}{The DMSS probe of the PMQ. Reproduced from ref. \cite{Volkwyn2008b}.}

\clearpage
\section{Codebook (to go in Supplemental Material)}
\begin{table*}
\caption{Codes for the RD probe \label{tab:RDcodeboook}}%
\begin{ruledtabular}
\begin{tabular}{p{1cm}p{1cm}p{1.2cm}p{3.8cm}p{8.5cm}}
\textbf{Probe} & \textbf{Number} & \textbf{Paradigm} & \textbf{Name}                                         & \textbf{Definition: ``Argument is that..."  }                                                                                                                                                                                                                    \\ \hline
RD    & P1     & P        & Measure the true value                       & ... the experimenter could measure the correct value in a single measurement.                                                                                                                                                                        \\ \hline
RD    & P2     & P        & Identify the outliers after all measurements & ...repeated measurements are needed in order to know which measurements were mistakes or outliers, after all measurements are taken. This code includes the idea that the experimenter must get the same result at least twice for it to be correct. \\ \hline
RD    & P3     & P        & Available time or resources                  & ... a course of action is better due to considerations about how much time or resources it would require                                                                                                                                                \\ \hline
RD    & P4     & P        & Need to practice as you go                   & ... practice is needed to account for errors/outside factors as measurements are being made                                                                                                                                                             \\ \hline
RD    & P5     & P        & Misc. point                                  & Point-like argument that doesn't fit the other point-like codes                                                                                                                                                                                      \\ \hline
RD    & S1     & S        & Measure a spread                             & ... multiple measurements will allow the experimenter to calculate/estimate a spread/variation/uncertainty                                                                                                                                              \\ \hline
RD    & S2     & S        & Measure an average                           & ... multiple measurements will allow the experimenter to calculate an average/mean                                                                                                                                                                      \\ \hline
RD    & S3     & S        & Use all the data together                    & ... multiple measurements will all be used together to improve accuracy/precision/goodness. Doesn't talk about average or spread specifically.                                                                                                          \\ \hline
RD    & S4     & S        & Reduce uncertainty of mean                   & ... multiple measurements will be used to reduce the error/uncertainty of the mean/average.                                                                                                                                                             \\ \hline
RD    & S5     & S        & Misc. set                                    & Set-like argument that doesn't fit the other set-like codes                                                                                                                                                                                          \\ \hline
RD    & U1     & U        & Just take more data                          & ... experimenter needs to take more data. No statistical reasoning apparent.                                                                                                                                                                            \\ \hline
RD    & U2     & U        & More data cancels out error                  & ... experimenter needs to take more data to cancel or outweigh the effect of error.                                                                                                                                                                     \\ \hline
RD    & U3     & U        & More data is better                          & ... more data is better / more accurate / more precise / etc. Includes if reasoning other than statistical reasoning apparent.                                                                                                                          \\ \hline
RD    & U4     & U        & Misc.                                        & Argument that doesn't fit into any of the other codes.                                                                                                                                                                                               \\ \hline
RD    & U5     & U        & Unintelligible                               & Unintelligible / blank / logically incoherent  
\end{tabular}
\end{ruledtabular}
\end{table*}

\begin{table*}
\caption{Codes for the UR probe \label{tab:URcodeboook}}%
\begin{ruledtabular}
\begin{tabular}{p{1cm}p{1cm}p{1.2cm}p{3.8cm}p{8.5cm}}
\textbf{Probe} & \textbf{Number} & \textbf{Paradigm} & \textbf{Name}                                         & \textbf{Definition: ``Argument is that..."  }                                                                                                                                                                                                        \\ \hline
UR    & P1     & P        & Choose single value                                  & ...experimenter should choose a single value to report (for any reason).                                                                                                                                                               \\ \hline
UR    & P2     & P        & Average as last resort                               & ...experimenter should report the average because no better option exists.                                                                                                                                                             \\ \hline
UR    & P3     & P        & Misc. point                                          & Point-like argument that doesn't fit the other point-like codes.                                                                                                                                                                       \\ \hline
UR    & S1     & S        & Simply “average”, or names reported value as average & States things like ``I averaged," ``do the average," ``average is best," or ``it is the average," but does not elaborate along the lines of the other codes. Includes statements that simply say what the reported value is.           \\ \hline
UR    & S2     & S        & Why average is useful                                & ...reporting the average is best, because (in general) it accounts for fluctuations or errors, or because it predicts future measurements.                                                                                             \\ \hline
UR    & S3     & S        & Why average is appropriate in this case              & ...reporting the average is best because all of this data matters, or because the spread of this data is small enough. Includes reporting all data as well as the average. Does not include ``it is the correct thing to do" (see S7). \\ \hline
UR    & S4     & S        & Report average and spread                            & ...experimenter should report the average and the uncertainty/range/spread.                                                                                                                                                            \\ \hline
UR    & S5     & S        & How to compute                                       & Response explains how to compute the average.  May be double coded when a separate explanation appears.                                                                                                                                \\ \hline
UR    & S6     & S        & Discard outliers, then average                       & ...experimenter should discard outliers/extreme data points, and then compute an average from the data that remains.                                                                                                                   \\ \hline
UR    & S7     & S        & Misc. set                                            & Set-like argument that doesn't fit the other set-like codes. Rule based reasons are coded here (e.g. ``logical thing to do" or ``the correct thing to do").                                                                            \\ \hline
UR    & U1     & U        & Misc.                                                & Argument that doesn't fit into any of the other codes.                                                                                                                                                                                 \\ \hline
UR    & U2     & U        & Unintelligible                                       & Unintelligible / blank / logically incoherent        
\end{tabular}
\end{ruledtabular}
\end{table*}

\begin{table*}
\caption{Codes for the SMDS probe \label{tab:SMDScodeboook}}%
\begin{ruledtabular}
\begin{tabular}{p{1cm}p{1cm}p{1.2cm}p{3.8cm}p{8.5cm}}
\textbf{Probe} & \textbf{Number} & \textbf{Paradigm} & \textbf{Name}                                         & \textbf{Definition: ``Argument is that..."  }    \\ \hline
SMDS &	P1	& P & 	The means are the same & 	...the groups agree because the means are the same. \\ \hline
SMDS &	P2 &	P &	Spreads don't matter &	...the fact that the spreads or individual trials are different does not matter, including responses that focus on agreement of the averages while providing a reason for why the sets are different. \\ \hline
SMDS &	P3 &	P &	A has fewer outliers &	...A is better because that group has fewer outliers, or A's individual measurements are more precise. Contains no reasoning about spread. \\ \hline
SDMS &	P4 &	P &	Differences in carefulness &	...differences in the spread are due to differences in how carefully the measurements were performed. \\ \hline
SMDS &	P5 &	P &	Chose B, no explanation	& Student chose ``B" but left the explanation blank. \\ \hline
SMDS &	P6 &	P &	Misc. point	& Point-like argument that doesn't fit the other point-like codes. \\ \hline
SMDS &	S1 &	S	& A is better &	...group A is better / more accurate / more precise / etc. No further explanation. \\ \hline
SMDS &	S2	& S	& Smaller spread is better, no mention of external factors	& ...a smaller spread/uncertainty/range is better / more accurate / more precise / etc. The response does not mention external factors, outliers, human error, etc. \\ \hline
SMDS &	S3 &	S	& Smaller spread is better, due to external factors	& ...a smaller spread/uncertainty/range is better / more accurate / more precise / etc. The response mentions external factors, outliers, human error, etc. \\ \hline
SMDS &	S4 &	S &	Chose A, no explanation	& Student chose ``A" but left the explanation blank. \\ \hline
SMDS &	S5 &	S	& Misc. set	& Set-like argument that doesn't fit the other set-like codes. \\ \hline
SMDS &	U1 &	U	 & Misc. &	Argument that doesn't fit into any of the other codes. \\ \hline
SMDS &	U2 &	U	& Unintelligible	& Unintelligible / blank / logically incoherent   
\end{tabular}
\end{ruledtabular}
\end{table*}

\begin{table*}
\caption{Codes for the DMSS probe \label{tab:DMSScodeboook}}%
\begin{ruledtabular}
\begin{tabular}{p{1cm}p{1cm}p{1.2cm}p{3.8cm}p{8.5cm}}
\textbf{Probe} & \textbf{Number} & \textbf{Paradigm} & \textbf{Name}                                         & \textbf{Definition: ``Argument is that..."  }    \\ \hline
DMSS &	P1 &	P	& Means and spreads must both match	& ...the groups do not agree because in order to agree, the means and the spreads must both match. \\ \hline
DMSS &	P2 &	P	& Means must match &	...the groups do not agree because the means are not the same (no mention of spread). \\ \hline
DMSS	& P3 &	P &	Means close enough, treats average as point &	...the groups agree because the means are close enough. \\ \hline
DMSS	& P4	& P	& Compare point-by-point, don't agree &	...the groups do not agree. Data are compared point by point. \\ \hline
DMSS &	P5 &	P &	Compare point-by-point, do agree &	...the groups agree. Data are compared point by point. \\ \hline
DMSS &	P6 &	P	& Chose B, blank explanation &	Student chose ``B" but left the explanation blank. \\ \hline
DMSS &	P7 &	P	& Misc. point &	Point-like argument that doesn't fit the other point-like codes. \\ \hline
DMSS &	S1 &	S &	Means are close enough, talks about statistical variation in general &	...the groups agree because the averages are close enough. Argument contains no reference to spreads, but does discuss statistical variation in general. \\ \hline
DMSS &	S2 &	S	& Similar means and spreads, no mention of overlap	& ...the groups agree because the means and spreads are similar. Argument does not consider the overlap between the means and/or spreads of the two datasets. \\ \hline
DMSS &	S3	& S	& Similar means and spreads, mentions overlap &	...the groups agree because the means and spreads are similar. Argument considers the overlap between the means and/or spreads of the two datasets. \\ \hline
DMSS &	S4	& S	& Chose A, blank explanation &	Student chose ``A" but left the explanation blank. \\ \hline
DMSS &	S5 &	S &	Misc. set &	Set-like argument that doesn't fit the other set-like codes. \\ \hline
DMSS &	U1 &	U	& Not about statistics &	...only non-statistical things, such as systematics, are mentioned. \\ \hline
DMSS &	U2 &	U	& Cannot calculate uncertainty/spread &	...the student states that they could not calculate the uncertainty or the spread, or that such values were not provided, with no further reasoning. \\ \hline
DMSS &	U3	& U &	Misc. &	Argument that doesn't fit into any of the other codes. \\ \hline
DMSS &	U4 &	U &	Unintelligible &	Unintelligible / blank / logically incoherent \\ \hline
\end{tabular}
\end{ruledtabular}
\end{table*}

\end{document}